\documentclass[12pt]{article}

\usepackage{amsmath,amsfonts,amssymb,verbatim,graphicx}

\newcommand{\bx}{\mathbf{x}}
\newcommand{\bg}{\mathbf{g}}

\newcommand{\bbf}{\mathbf{f}}
\newcommand{\bu}{\mathbf{u}}
\newcommand{\bv}{\mathbf{v}}
\newcommand{\bw}{\mathbf{w}}

\newcommand{\bnabla}{\boldsymbol{\nabla}}

\newcommand{\rd}{{\text{\rm d}}}
\newcommand{\ddt}[1]{\frac{\text{\rm d}#1}{\text{\rm dt}}}
\newcommand{\supp}{\operatorname*{supp}}
\newcommand{\Vinner}[1]{(\!({#1})\!)}

\newcommand{\be}{\begin{equation}}
\newcommand{\ee}{\end{equation}}
\newcommand{\la}{\label}
\newcommand{\ba}{\begin{array}{c}}
\newcommand{\ea}{\end{array}}

\newtheorem{thm}{Theorem}
\newtheorem{prop}{Proposition}

\newtheorem{lemma}{Lemma}
\newtheorem{definition}{Definition}
\newtheorem{cor}{Corollary}

\title{Invariant measures for the $3$D Navier-Stokes-Voigt equations and their Navier-Stokes limit}

\author{Fabio Ramos${}^{1,2}$
\quad Edriss S. Titi${}^{1,3}$ \vspace{1cm}\\
${}^1$ Department of Computer Science and Applied Mathematics \\
The Weizmann Institute of Science  \\
Rehovot 76100, Israel\\
edriss.titi@weizmann.ac.il\\
${}^2$ Institute of Applied Mathematics,
University of Bonn \\
Bonn 53115, Germany\\
ramos@iam.uni-bonn.de\\
\smallskip
${}^3$ Department of Mathematics \\
and Department of Mechanical and  Aerospace Engineering \\
University of California \\
Irvine, CA  92697-3875, USA\\
etiti@math.uci.edu\\
\mbox{ }\\
\centerline{This work is dedicated to Professor Roger Temam}\\ \centerline{on the occasion of his seventieth birthday}
}

\date{October 4, 2009}

\begin{document}

\maketitle

\noindent{\bf Abstract.} The Navier-Stokes-Voigt model of
viscoelastic incompressible fluid has been recently proposed as a
regularization of the three-dimensional Navier-Stokes equations for
the purpose of direct numerical simulations. Besides the kinematic
viscosity parameter, $\nu>0$, this model possesses a regularizing
parameter, $\alpha> 0$, a given length scale parameter, so that
$\frac{\alpha^2}{\nu}$ is the relaxation time of the viscoelastic
fluid.  In this work, we derive several statistical properties of
the invariant measures associated with the solutions of the
three-dimensional Navier-Stokes-Voigt equations. Moreover, we prove
that, for fixed viscosity , $\nu>0$, as the regularizing parameter
$\alpha$ tends to zero, there exists a subsequence of probability
invariant measures converging, in a suitable sense, to a strong
stationary statistical solution of the three-dimensional
Navier-Stokes equations, which is a regularized version of the
notion of stationary statistical solutions - a generalization of the
concept of invariant measure introduced and investigated by Foias.
This fact supports earlier numerical observations, and provides an
additional evidence that, for small values of the regularization
parameter $\alpha$, the Navier-Stokes-Voigt model can indeed be
considered as a model to study the statistical properties of the
three-dimensional Navier-Stokes equations and turbulent flows via
direct numerical simulations.

\vskip 0.25in

\noindent{\bf MSC:} 76D06,76D05,76F20,76F55,76A10

\vskip 0.25in

\noindent{\bf Keywords:} Navier-Stokes-Voigt, Navier-Stokes-Voight,
inviscid regularization, Statistical solutions, turbulence,
viscoelastic flows.

\section{Introduction}

In this work, we consider the three-dimensional Navier-Stokes-Voigt
equations (NSV), a viscoelastic incompressible fluid model
introduced by Oskolkov in \cite{oskolkov1}, and which was proposed
by the authors of \cite{ctk} as a smooth inviscid regularization of
the $3$D Navier-Stokes equations (NSE) for the purpose of direct
numerical simulations (DNS). More specifically, we consider the NSV
model subject to periodic or no-slip boundary conditions, and driven
by a given force field $\mathbf{f}$. The velocity vector field,
$\bv(\bx,t)$, and the scalar kinematic pressure, $p(\bx,t)$, are
governed by the system of equations \be \left \{ \ba
\partial_t (\bv - \alpha^2\Delta \bv) -\nu \Delta \bv +\bv\cdot\nabla \bv   + \nabla p = \mathbf{f},\quad \bx\in\Omega,\\
\nabla\cdot \bv = 0,\quad \bx\in\Omega,\\
\bv(\bx,0)=\bv_0(\bx),\quad \bx\in\Omega,\\
\bv(\bx,t)=0\quad
\bx\in\partial\Omega,\quad\text{or}\quad\bv(\bx,t)\;\text{is}\;\text{periodic};
\ea \right. \la{aueq} \ee in the smooth domain $\Omega\subset
\mathbb{R}^3$, in the case of no-slip Dirichlet boundary condition,
or with basic periodic domain $\Omega = [0, L]^3\subset
\mathbb{R}^3$, when equipped  with periodic boundary conditions.
Here, $\alpha\geq 0$ is a given length scale parameter, and $\nu>0$
is a given kinematic viscosity, such that $\alpha^2/\nu$ is the
relaxation time of the viscoelastic fluid.

 In the periodic case, we assume that the driving force, and the initial velocity $\bu_0$ satisfy
 \[
\int_\Omega\bu_0(\bx)\;d\bx=\int_\Omega\mathbf{f}(\bx)\;d\bx=0,
\]
and it is easy to see that it implies
\[
\int_\Omega\bv(\bx,t)\;d\bx=0,\quad\forall t\geq 0.
\]

The Navier-Stokes-Voigt model of viscoelastic incompressible fluid,
\eqref{aueq}, (sometimes written as Navier-Stokes-Voight) was
introduced by Oskolkov in \cite{oskolkov1}, and pointed out by O.
Ladyzhenskaya as one of the reasonable modifications of the
Navier-Stokes equations, see \cite{ladyosk}. In \cite{oskolkov1}, A.
P. Oskolkov studied and proved its solvability in different
functional spaces. We  remark that these equations behave like a
damped hyperbolic (pseudo-parabolic) system, see \cite{klt}, and,
therefore, the solutions do not experience fast (instantaneous)
smoothening of the initial data, as it is for parabolic systems like
the Navier-Stokes equations. This fact could prevent the NSV model
from being a reasonable modification of the Navier-Stokes equations,
however, since we are proposing it as a model for direct numerical
simulations of turbulent flows in statistical equilibrium, i.e.,
after the solutions reach the global attractor, we are mainly
interested in its long time behavior. Indeed,  it was proved in
\cite{KLT}, that solutions in the global attractor are smooth, if
the forcing field is  smooth enough, even for initial data
satisfying only finite kinetic energy and finite enstrophy (i.e.
bounded in the Sobolev $H^1$-norm). In particular, in \cite{KLT}, it
is shown in the periodic case, that if the forcing field is
analytic, then the global attractor consists of analytic functions.
This result, in   conjunction with results proved in this work,
proves that if the forcing field is smooth enough, then averaged
structure functions, with respect to an invariant measure for the
NSV flow, display exponential decaying tail.

As it was observed above, the NSV model presents an extra length scale associated with the
 viscoelastic properties of the fluid, the parameter $\alpha$, besides the well known
 Kolmogorov length scale, $\eta$, (see, e.g., \cite{frisch} for its definition), which is
 usually associated with the smallest scales of motion in turbulent flows.
In \cite{LRT}, based on numerical simulations of the Sabra Shell
model, it is observed that for a large range of values of
$\alpha>0$, i.e. $\alpha\gg\eta$, there are two distinct regions
associated with the inertial range of the energy spectrum for the
NSV model. The first one obeying the celebrated Kolmogorov $
k^{-2/3}$ power law (with anomalous correction), followed by a
second range of wavenumbers, where energy condensates, and it is
simply equipartitioned. The range of the second power-law, however,
slowly disappears as $\alpha/\eta$  decreases, restoring the usual
Navier-Stokes inertial range regime, when $\alpha \ll \eta  $.

 The numerical simulations of the Shell model presented in \cite{LRT} also suggest that for a large range
 of values of $\alpha>0$, when $\alpha\gg\eta$,   small scales velocity fluctuations are vigorously damped,
 due to a slowness of the energy transfer timescales associated with these small length scales. This has an
 extraordinary effect in reducing the energy dissipation rate intermittency, characterized by violent
 fluctuations away from the average in the energy dissipation rate.
It is therefore suggested in \cite{LRT} that by tuning the parameter
$\alpha$, one may attenuate the strong
 velocity fluctuations related to the intermittent events, reducing thus the stiffness of DNS of turbulent flows,
 with only a small effect on the energy containing scales. This reduction of stiffness in the NSV model has also
 been observed in numerical experiments in implementing the NSV
 model for inpainting \cite{EHL}.

The numerical experiments in \cite{LRT} were performed using the Sabra shell phenomenological model
of turbulent flows. However, because these models do not present any spatial structure, it is not clear
if such properties persist for full direct numerical simulations of $3$D turbulent flows. In fact,
a large class of viscoelastic flows present anomalous behavior even for very small viscoelastic parameters,
see, e.g., \cite{lumley}, and such phenomenon could be present in flows governed by these equations.

In this work, we rigorously establish several statistical properties
of the energy spectrum of the $3$D NSV model that were observed in
the shell model simulations in \cite{LRT}, using the notion of
invariant measures. For example, by considering the results of
\cite{KLT}, we can rigorously justify the exponentially decaying
dissipation range observed in the simulations presented in
\cite{LRT}, despite the fact that these equations behave like a
damped hyperbolic (pseudo-parabolic) system \cite{klt}.

 Concerning our main result, we start by defining a notion of strong stationary statistical solution
 of the Navier-Stokes equations, which is a regularized version of the concept of stationary statistical
 solution that was introduced and investigated by Foias in \cite{foias72}, \cite{foias73} (see also \cite{fmrt}).
 This notion is a generalization of the concept of an invariant measure for the semigroup generated
 by the three-dimensional Navier-Stokes equations.

 Our main result states the following: Fix $\nu>0$, and let $\mu^{\alpha_n}$ be a given  sequence of
 invariant measures for the semigroups generated by the NSV model. Then, there exists a subsequence,
 $\mu^{\alpha_{n_{j}}}$, weakly converging to a strong
 stationary statistical solution  of the
 Navier-Stokes equations, as $\alpha_{n_{j}}\to 0$.

This fact supports the observations reported in \cite{LRT},
providing a rigorous evidence that, for  small values of the
regularization parameter $\alpha$, the NSV model approximates
several statistical properties of the $3$D Navier-Stokes equations,
and, therefore, may be used as a reliable subgrid scale model for
direct numerical simulations of turbulent flows.

\section{Mathematical framework}

We will follow the standard functional formulation for the Navier-Stokes equations, see, e.g., \cite{constantinfoias, fmrt, temam}, and the functional formulation for the NSV equations used in \cite{klt}. We denote by $\mathcal{V}$, in the no-slip case, the set
\be
\label{smoothf}
\mathcal{V}=\left\{\varphi\in C_0^\infty(\Omega);\;\nabla\cdot\varphi=0\right\}.
\ee
In the periodic case, we define $\mathcal{V}$ as
\be
\label{persmoothf}
\begin{aligned}
 \mathcal{V}=&\left\{\mbox{All the three-dimensional vector valued trigonometric polynomials $\varphi$}\right.\\
&\left. \mbox{with basic periodic domain $\Omega=[0,L]^3$; $\nabla\cdot\varphi=0$ and $\int_{\Omega}\varphi\;dx=0$}\right\}.
\end{aligned}
\ee
The two fundamental functional spaces in this work are defined by
\[
H= \text{closure of } \mathcal{V} \text{ in } \left(L^2(\Omega)\right)^3,
\]
and
\be
\la{vdefin}
 V= \text{closure of } \mathcal{V} \text{ in } \left(H^1(\Omega)\right)^3.
 \ee
The inner products in $H$ and $V$ are denoted, respectively, by
\[ (\bu,\bv)= \int_\Omega \bu(\bx)\cdot\bv(\bx) \;\rd\bx,
     \quad \Vinner{\bu,\bv} = \int_\Omega \sum_{i=1}^{3}
        \frac{\partial \bu}{\partial x_i}
          \cdot \frac{\partial \bv}{\partial x_i}\; \rd\bx,
\]
and the associated norms by $\left|\bu\right|=(\bu,\bu)^{1/2}$,
$\|\bu\|=\Vinner{\bu,\bu}^{1/2}$ (the latter is a norm thanks to the
Poincar\'e inequality~(\ref{Poinc}), below).

We denote by $P_{\text{LH}}$ the (Leray-Helmholtz) orthogonal
projector in $L^2(\Omega)^3$ onto the subspace $H$. The Stokes operator is defined
by $A\bu = - P_{\text{LH}}\Delta \bu$, from $D(A)=H^2\cap V$ into $H$. The norm in $D(A)$ is
$$\left\|\bu\right\|_{D(A)}=|A\bu|,\quad\forall\bu\in D(A).$$
Moreover, because the inverse of the Stokes operator is a positive self-adjoint compact operator in $H$, we can define its powers $A^s$, $s \in \mathbb{R}$, with domain $D(A^s )$.
We have $V = D(A^{1/2})$ and its dual $V' = D(A^{-1/2})$, see, e.g., \cite{constantinfoias}, \cite{fmrt}, or \cite{temam}.

The term $B(\bu,\bv) = P_{\text{LH}}((\bu\cdot\bnabla)\bv)$ is a bilinear term
associated with the inertial term, which satisfies
\be
\la{bilinear}
B(\bu,\bv)\in V',\quad \bu,\bv\in V,
\ee
where $V'$ denotes the dual space of $V$, (see, e.g., \cite{constantinfoias, temam}). Thus, taking the duality action in $V\times V'$, between the bilinear term, $B(\bu,\bv)$ and a
vector field $\bw\in V$ yields a trilinear term (see, e.g., \cite{constantinfoias, temam}),
\[
 b(\bu,\bv,\bw) = \langle B(\bu,\bv),\bw\rangle_{V',V},
\]
which is defined for $\bu,\bv, \bw\in V$, where $\langle\cdot,\cdot\rangle_{V,V'}$ denotes the duality action between the spaces $V$, and its dual space $V'$. An important relation for
the trilinear term is the orthogonality property, see, e.g., \cite{constantinfoias, temam},
\begin{equation}
b(\bu,\bv,\bv) = 0,
\end{equation}
for $\bu,\bv\in V$. It follows from this relation the anti-symmetry property (see, e.g., \cite{constantinfoias, temam}),
\begin{equation}
\label{asym}
b(\bu,\bv,\bw) = - b(\bu,\bw,\bv),
\end{equation}
for all $\bu,\bv,\bw\in V$.

Moreover, because the inverse of the Stokes operator is a positive self-adjoint compact operator in $H$, then there exists  a complete orthonormal basis of $H$ formed by eigenvectors, $\{\mathbf{w}_{j}\}_{j}$, with associated eigenvalues, $A\mathbf{w}_j=\lambda_j\mathbf{w}_j$, satisfying $0<\lambda_1\leq\lambda_2\leq\ldots\leq\lambda_{j}\rightarrow\infty$, as $j\rightarrow\infty$, see, e.g.,  \cite{constantinfoias, fmrt, temam} for details.  In this geometry and setting, the Poincar\'e inequality holds
\begin{equation}
\la{Poinc}
  \left|\bv\right|^2\leq \lambda_{1}^{-1}\left\|\bv\right\|^2,
\end{equation}
where $\lambda_{1}$ is the first eigenvalue of the
Stokes operator.

We will need the following inequalities, which are consequences of H\"older inequality, Poincar\'e inequality, and well-known inequalities of Ladyzhenskaya, see, e.g., \cite{constantinfoias}, \cite{fmrt}, or \cite{temam} :
\be
\la{b5}
|b(\bu,\bv,\bw)|\leq C|\bu|\left\|\bv\right\| |\bw|^{1/4} |A\bw|^{3/4} \leq C|\bu||A\bv||A\bw|,
\ee
for every $\bu\in H$, and $\bv,\bw\in D(A)$. Also,
\be
\la{b6}
|b(\bu,\bv,\bw)|\leq C|\bu|^{1/4}\left|A\bu\right|^{3/4}\left\|\bv\right\| |\bw|\leq C|A\bu||A\bv||\bw|,
\ee
for every $\bu, \bv\in D(A)$, and $\bw\in H$.

Throughout this work, if not otherwise stated, we assume a forcing field $\mathbf{f}\in H$. If we apply the Leray-Helmholtz operator, $P_{LH}$, to the first equation of the system \eqref{aueq}, we obtain the following equivalent functional differential equation
\begin{equation}
  \label{fnseeq}
  \ddt{\bv} + \alpha^2\ddt{A\bv} = \mathbf{F}(\bv) :=(\mathbf{f} - \nu A\bv - B(\bv,\bv)),\quad\;
  \bv(0)=\bv_0.
\end{equation}
 Global existence and uniqueness of \eqref{fnseeq} was first
 studied in \cite{oskolkov1}, where it was established that
 the system \eqref{aueq} generates a continuous
 semigroup, $S^{\alpha}(t):V\to V$, $t\in\mathbb{R}^{+}$, i.e.,
 a one-parameter family of maps $\left\{S^\alpha(t)\right\}_{t\geq 0}$,
 satisfying the properties of a continuous semigroup, such that
 the solution of \eqref{aueq} (or equivalently, of \eqref{fnseeq}),
 satisfies $\bv(t)=S(t)\bv_0$, for every $\bv_0\in V$.
 In \cite{ctk}, global regularity of \eqref{fnseeq} was
 also proved for the inviscid model, i.e. when $\nu=0$. Moreover,
 higher-order regularity of the inviscid case, i.e. the Euler-Voigt
 model, is established in \cite{Larios-Titi}; furthermore, a new
 blowup criterion for the three-dimensional Euler equations, by
 means of this inviscid regularization is introduced.

\section{Energy budget for the NSV model}
The global regularity result for solutions of the Navier-Stokes-Voigt equations established in $\cite{oskolkov1}$ implies that the following energy equality holds for every $t\in \left[0,\infty\right)$:
\begin{equation}
\label{enequation} \frac{d}{dt}(\frac{1}{2}\left|\bv(\cdot,
t)\right|^2+\frac{\alpha^2}{2}\left\|\bv(\cdot,
t)\right\|^2)=(\mathbf{f},\bv(\cdot, t))-\nu\left\|\bv(\cdot,
t)\right\|^2,
\end{equation}
(in fact the energy equality and global well-posedness holds for all
$t\in \mathbb{R}$). Similarly, the global regularity results
established in \cite{ctk} imply that the solutions of the NSV
equations in the inviscid (i.e. Euler-Voigt model, $\nu=0$)  and
unforced setting, $\mathbf{f}=0$, satisfy for every $t\in
\mathbb{R}$:
\begin{equation}
\label{invenequation}
\frac{d}{dt}(\frac{1}{2}\left|\bv^0(\cdot, t)\right|^2+\frac{\alpha^2}{2}\left\|\bv^0(\cdot, t)\right\|^2)=0.
\end{equation}
Therefore, the conserved quantity in the inviscid and unforced
setting of the NSV (i.e. Euler-Voigt) model is
\begin{equation}
\mathcal{E}_{\alpha}=\frac{1}{2}\left|\bv\right|^2+\frac{\alpha^2}{2}\left\|\bv\right\|^2,
\end{equation}
which we call the $\alpha$-energy. The quantity
\begin{equation}
\mathcal{K}_{\alpha}=\frac{1}{2}\left|\bv\right|^2
\end{equation}
is the usual kinetic energy. We remark that while the kinetic energy is formally conserved for the inviscid and unforced incompressible Navier-Stokes equations, this is not the case for the inviscid and unforced setting of the NSV model. Those differences in the conserved quantity for the inviscid and unforced case are the main source of deviation between the statististical properties of the NSV model and of the Navier-Stokes equations observed in \cite{LRT}. In Section $6$, we establish a result concerning the convergence of the averaged kinetic energy of the NSV model as the parameter $\alpha$ tends to zero.

Let $\mathbf{w}_j$ be the orthonormal basis of $H$ composed by the eigenfunctions of the Stokes operator, $A$. For a vector field $\bv\in H$,  we define the component $\bv_\kappa$, for a wavenumber $\kappa$, by
\be
\bv_{\kappa}=\sum_{\lambda_j=\kappa^2}(\bv,\mathbf{w}_j)\mathbf{w}_j.
\la{proj1}
\ee
We also define the component $\bv_{\kappa',\kappa''}$ by
\be
\bv_{\kappa',\kappa''}=\sum_{\kappa'\leq\kappa<\kappa''}\bv_\kappa.
\la{proj2}
\ee
Then, we can write the projected Navier-Stokes-Voigt equations on the shell $[\kappa',\kappa'')$
\be
\label{projected}
\frac{d}{dt}(\bv_{\kappa',\kappa''}+\alpha^2 A\bv_{\kappa',\kappa''})+\nu A\bv_{\kappa',\kappa''}+B(\bv,\bv)_{\kappa',\kappa''}=\mathbf{f}_{\kappa',\kappa''}.
\ee

Let us now obtain the $\alpha$-energy budget for the shell $[\kappa',\kappa'')$. Multiplying in $L^2$ by $\bv_{\kappa',\kappa''}$, we obtain
\be
\begin{aligned}
&\frac{1}{2}\frac{d}{dt}(\left|\bv_{\kappa',\kappa''}\right|^2+\alpha^2\left\|\bv_{\kappa',\kappa''}\right\|^2)+\nu\left\|\bv_{\kappa',\kappa''}\right\|^2=b(\bv,\bv,\bv_{\kappa_1,\kappa''})\\
&-b(\bv,\bv,\bv_{\kappa_1,\kappa'})
+(\mathbf{f},\bv_{\kappa',\kappa''})=[e^{\alpha}_{\kappa'}(\bv)-e^{\alpha}_{\kappa''}(\bv)]+(\mathbf{f},\bv_{\kappa',\kappa''}),
\la{tbud3}
\end{aligned}
\ee
where
\[
e^{\alpha}_\kappa(\bv):=e_{\kappa}^{\alpha,\rightarrow}(\bv)-e^{\alpha,\leftarrow}_{\kappa}(\bv)
\]
is the net rate of $\alpha$-energy transfer at $\kappa$, and
\[
e_{\kappa}^{\alpha,\rightarrow}(\bv):=-(B(\bv_{\kappa_1,\kappa},\bv_{\kappa_1,\kappa}),\bv_{\kappa,\infty})
\]
represents the net rate of $\alpha$-energy from the lower modes to the higher modes, and
\[
e_{\kappa}^{\alpha,\leftarrow}(\bv):=-(B(\bv_{\kappa,\infty},\bv_{\kappa,\infty}),\bv_{\kappa_1,\kappa})
\]
represents the net rate of $\alpha$-energy from the higher modes to the lower modes.

\noindent{\bf{Remark.}}
 Since we are dealing simultaneously with $\alpha>0$
 and $\alpha=0$, then for $\kappa''=\infty$, all the calculations
 above are formal. However, they can all be rigorously recovered
 for the NSV case, i.e. when $\alpha>0$, via the Galerkin
 approximation procedure. For the Navier-Stokes case, i.e.
 when $\alpha=0$, if $\kappa''=\infty$, we can only state
 equations \eqref{projected} and \eqref{tbud3}  in terms of
 inequalities, due to a possible lack of regularity for the
 three-dimensional Navier-Stokes equations,
 see, e.g., \cite{fmrt}.

\section{Invariant measures and stationary statistical
solutions for the NSV model}

A rigorous mathematical framework for investigating the statistical
properties of turbulent fluid flows in statistical equilibrium was
first considered by Hopf in \cite{hopf}. In \cite{foias72, foias73},
Foias established the notion of stationary statistical solutions of
the Navier-Stokes equations, which is a generalization of the
concept of invariant probability measures for the semigroup
associated with the solutions of the equations of motion. This
notion is important because the semigroup generated by the
Navier-Stokes equations is not known to be well-posed in $V$, and,
therefore, the notion of invariant measure in $V$ is not well-posed
either. A detailed discussion of this issue can be found in
\cite{fmrt}.

Because of the global regularity of solutions of the NSV model in the space $V$ obtained in \cite{oskolkov1}, one can refrain from using the abstract notion of stationary statistical solutions, and work only with the more familiar notion of invariant measure.

However, because the main goal of this work is to show convergence results of the statistical properties of the NSV model to the corresponding ones of the Navier-Stokes equations, as $\alpha\to 0$, we will show first that invariant measures of the NSV model are stationary statistical solutions of the NSV model, and using this fact, we prove a weak convergence theorem in the sense of stationary statistical solutions of the Navier-Stokes equations. This approximation result is the main reason why we are interested in deriving several statistical properties of the invariant measures associated with the NSV model in this section.

 We recall that a probability measure $\mu^{\alpha}$ on $V$ is called an invariant measure for the semigroup $\left\{S^{\alpha}(t)\right\}_{t\geq 0}$ if
 \[
 \mu(E)=\mu((S^{\alpha}(t))^{-1}E), \quad\forall t\geq 0,
 \]
for every Borel measurable set $E$ in $V$. This definition is equivalent to say that
\[
\int_{V}\Phi(\bu)d\mu^{\alpha}(\bu)=\int_{V}\Phi(S^{\alpha}(t)\bu)d\mu^{\alpha}(\bu), \quad\forall \Phi\in L^1(\mu).
\]
 It is easy to see that the Dirac measure concentrated at the
 steady state solutions of the NSV model, which coincide with
 steady solutions of the Navier-Stokes equations,  are invariant
 measures for the semigroup
 $\left\{S^{\alpha}(t)\right\}_{t\geq 0}$ generated by the NSV
 equations. Therefore, because the set of steady states
 is nonempty (see, e.g., \cite{constantinfoias}, \cite{Foias-Temam}
 and \cite{temam}), we have that the set of invariant
 measures for $\left\{S^{\alpha}(t)\right\}_{t\geq 0}$ is nonempty.

We recall that the support of an invariant measure, $\mu^{\alpha}$,  consists only of points that are nonwandering, i.e., for any $\bu\in\supp \mu^{\alpha}$, and any Borel set $E$ containing $\bu$, there exists a sequence of positive times $t_n\to \infty$, such that $S^{\alpha}(t_n)E\cap E\neq \emptyset$, for all $n\in \mathbb{Z}^+$, see, e.g., \cite{fmrt}. This implies that the support of an invariant measure, $\mu^{\alpha}$, is included in the global attractor of the semigroup $S^{\alpha}(t)$.

 In \cite{klt}, it was proved that $S^{\alpha}(t):V\to V$ has an absorbing ball in $V$, and it is an asymptotically compact semigroup, implying the existence of a global attractor in $V$. Moreover, the global compact attractor was shown to be bounded in $D(A)$ and to have finite Hausdorff and fractal dimensions.

  Furthermore, despite the fact that the NSV equations behave like a
  damped hyperbolic (pseudo-parabolic) system, rather than a parabolic equation, it was proved in \cite{KLT} that the $3$D periodic NSV equations possess an asymptotic smoothing property. More specifically, let us define the Gevrey class of functions
  \be
  \la{gevreydef}
  G^r_\tau :=D(A^{r/2}e^{\tau A^{1/2}})=\left\{\bu\in H;\; |A^{r/2} e^{\tau A^{1/2}}\bu|^2=\sum_{j \in \mathbb{Z}^3\setminus\left\{0\right\}}|\hat\bu_j|^2 |j|^{2r}e^{2\tau |j|}<\infty \right\},
  \ee
  where $\hat\bu_j$ are the correspondent Fourier coefficients of the $\Omega$-periodic vector field $\bu$, $\tau >0$, and $r\geq 0$. The space is equipped with the corresponding inner product
  \be
  (\bu,\bv)_{r,\tau}=(A^{r/2}e^{\tau A^{1/2}},A^{r/2}e^{\tau A^{1/2}})=\sum_{j\in\mathbb{Z}^3\setminus\left\{0\right\}} \hat\bu_j \cdot \hat\bv_j|j|^{2r}e^{2\tau |j|}<\infty,
  \ee
 for $\bu,\bv\in G_\tau^r$, and with the corresponding norm
  \be
  |\bu|_{r,\tau}=|A^{r/2}e^{\tau A^{1/2}}\bu|,
  \ee
  for $\bu\in G_\tau^r$.  One can prove that the
  space of real analytic functions $C^\omega(\Omega)$ has
  the following characterization:
  \[
  C^\omega(\Omega)=\bigcup_{\tau>0}G^r_\tau,
  \]
  for any $r\geq 0$, see, e.g., \cite{doelmantiti}, \cite{levermore} for details.
   In \cite{KLT}, it was proved that if the driving force, $\mathbf{f}$, belongs to a Gevrey class of functions, $G_{\tau_0}^1$,  for some $\tau_0>0$, then for every solution $\bv(\bx,t)$ of the $3$D periodic NSV equations, with initial data $\bv_0\in V$, there exists $t_0>0$, and a function $\bw(t)\in L^{\infty}(0,\infty, G_{\tau}^2)$, for some $\tau >0$ depending only on $|f|_{1,\tau_0}$, $\nu$, $\lambda_1$ and $\alpha$, so that
  \[
  \lim_{t\to\infty}\left\|\bu(t)-\bw(t)\right\|=0.
  \]
This result proves, in particular, that if $\mathbf{f}$ is analytic,
then the global attractor consists of analytic functions. We recall
that this technique of Gevrey class regularity was first introduced
by Foias and Temam \cite{FT89_Gevrey} for proving the analyticity,
in space and time, for strong solutions of the three-dimensional
Navier-Stokes equations, for short time (see also
\cite{Ferrari-Titi} for generalization of the technique to parabolic
analytic equations).

The following theorem is a straightforward application of the facts discussed above:

\begin{thm}
\label{testethm}
Let $\mu^\alpha$ be an invariant measure for the semigroup $\left\{S^{\alpha}(t)\right\}_{t\geq 0}$ generated by the $3$D periodic NSV equations, \eqref{aueq}. If the forcing field $\mathbf{f}$ belongs to $H$, then the $\supp\mu^{\alpha}$ is a subset of the global attractor, $\mathcal{A}$, of the semigroup generated by the $3$D NSV model, \eqref{aueq}, which is bounded in  $D(A)$. Moreover, if $\mathbf{f}\in G_{\tau_0}^1$,  for some $\tau_0>0$, then $\supp\mu^{\alpha}\subset \mathcal{A}\subset G_{\tau}^2 $, for some $\tau>0$, which implies
\[
\int_{V}\left\|e^{\tau A^{1/2}}\bu\right\|^2d\mu^{\alpha}(\bu)=\int_{V}\sum_{\kappa\in\mathbb{Z}^3\setminus\left\{0\right\}}\left|\bu_\kappa\right|^2|k|^2 e^{2\tau |\kappa|}d\mu^{\alpha}(\bu)<\infty.
\]
\end{thm}

\noindent{\bf{Remark.}} We remark that the trigonometric
polynomials, usually considered as forcing fields in the numerical
investigations of turbulent flows, belong to the class of analytic
functions. Therefore, our result applies to a wide set of
simulations. In particular, it justifies the observed exponential
tail in Sabra shell model simulations of the NSV equations in
\cite{LRT} (see similar results concerning the three-dimensional
Navier-Stokes equations \cite{DT95_Decay}).

\medskip

  Now, we define the notion of stationary statistical solutions
  of the Navier-Stokes-Voigt equations inspired by Foias
  \cite{foias72, foias73}, and \cite{fmrt}.

\begin{definition}
\label{nsvort}

A stationary statistical solution of the $3$D Navier-Stokes-Voigt model, \eqref{aueq}, is a Borel probability measure $\mu^{\alpha}$ on $V$ such that
\begin{enumerate}

\item[(1)]
\label{nsv1}
$\displaystyle{\int_{V}\left\|\bu\right\|_{D(A)}^{2}d\mu^{\alpha}(\bu)<\infty;}$

\item[(2)]
\label{nsv2}
$\displaystyle{\int_{V}( (I+\alpha^2A)^{-1}\Psi'(\bu),\mathbf{f}-\nu A\bu -B(\bu,\bu)) d\mu^{\alpha}(\bu)=0,}
\\
      \text{for any test functional}\; \Psi\in\mathcal{T}^\alpha, \text{where $\mathcal{T}^\alpha$ is defined in Definition } \ref{testealpha}.$

\item[(3)]
\label{nsv3}
$
\displaystyle{\int_{ \left\{E_1\leq \left| \bu\right|^2+\alpha^2\left\| \bu\right\|^2< E_2\right\}}\left(\nu\left\| \bu\right\|^2-( \mathbf{f}, \bu)\right)d\mu^{\alpha}(\bu)= 0}, \quad \mbox{for every }\, 0 \leq E_1 < E_2<\infty.$

\end{enumerate}
\end{definition}

\begin{definition}
\label{testealpha}
We define the class $\mathcal{T}^\alpha$ of test functions to be the set
of all real-valued functionals $\Psi=\Psi(\bu)$ on $H$ that are
bounded on bounded subsets of $V$ and such that the following
conditions hold:
\begin{enumerate}
\item
For any $\bu\in D(A)$, the Fr\'echet derivative $\Psi'(\bu)$ taken in
$H$ along vectors in $D(A)$ exists. More precisely, for each $\bu\in D(A)$, there
exists an element in $H$ denoted by $\Psi'(\bu)$ such that
\begin{equation}
\frac{\left|\Psi(\bu+\bv)-\Psi(\bu)-(\Psi'(\bu),\bv)\right|}{\left|\bv\right|}\rightarrow
0\quad\text{as}\; \left|\bv\right|\rightarrow 0, \bv \in D(A).
\end{equation}
\item
$\Psi'(\bu)\in H$ for all $\bu\in D(A)$, and the map $\bu\rightarrow
\Psi'(\bu)$ is continuous and bounded as a function from $H$ into
$H$.
\end{enumerate}
\end{definition}

 For example, we can take the cylindrical test functions
$\Psi:H\rightarrow\mathbb{R}$ of the form
$\Psi(\bu)=\psi\left((\bu,\mathbf{g}_{1}),\ldots,(\bu,\mathbf{g}_{m})\right)$,
where $\psi$ is a $C^{1}(\mathbf{R}^m;\mathbf{R})$ scalar function on $\mathbb{R}^{m}$,
$m\in\mathbb{N}$, with compact support, and
$\mathbf{g}_{1},\ldots,\mathbf{g}_{m}$ belong to $H$. For this
case we have
\[
\Psi'(\bu)=\sum_{j=1}^{m}\partial_{j}\psi((\bu,\mathbf{g}_{1}),\ldots,(\bu,\mathbf{g}_{m}))\mathbf{g}_{j},
\]
where $\partial_{j}\psi$ denotes the derivative of $\psi$ with
respect to the $j$-th variable. In this case, it follows that $\Psi '(\bu)\in H$
since it is a linear combination of the $\mathbf{g}_{j}$.

The class of test functions $\mathcal{T}^\alpha$ are broader than the class considered in the usual definition of
stationary statistical solutions of the Navier-Stokes equations, see, e.g., \cite{fmrt}. This is made possible by the
fact that solutions of the NSV model \eqref{aueq} are globally regular in $V$, see \cite{ctk, oskolkov1}.

 Condition $1$ in Definition \ref{nsvort}  implies that the support of the stationary statistical solutions are
included in $D(A)$. Condition $3$ is a local energy balance
equation, which implies that the stationary statistical solutions
have supports which are bounded in $V$, see Proposition
\ref{boundsupprop} below. Moreover, in Corollary
\ref{boundsupp6cor}, we show that for fixed $\alpha_0>0$, Condition
3 implies that the supports are uniformly bounded $H$, for all
$\alpha\in (0,\alpha_0]$.

 We also remark that because we consider weakly converging subsequences of probability measures in the Section $6$,
 we work with the weak topology of $V$. However, because $V$ is a separable Hilbert space, the Borel $\sigma$-algebra
generated by the weakly open sets coincides with that for the
open sets in the strong topology of $V$, see, e.g., \cite{DS}.
Thus, we identify these two probability spaces in the rest of
the work.

 Now, we prove a result concerning bounds for the $\supp\mu^\alpha$ in the space $V$. This result will be important because we will derive from it a uniform bound, with respect to $\alpha$, in the $H$-norm.

\begin{prop}
\label{boundsupprop}
Let $\mu^\alpha$ be a stationary statistical solution of the Navier-Stokes-Voigt equations. Then,
\be
\label{boundsupp}
\supp\mu^{\alpha}\subset K_{\alpha}:=\left\{ \bu\in V; \left| \bu\right|^2+\alpha^2\left\| \bu\right\|^2< (\lambda_1^{-1}+\alpha^2)\frac{\left|\mathbf{f}\right|^2}{\nu^2}\right\}.
\ee
\end{prop}
\noindent{\bf{Proof.}}
We follow the arguments used in \cite{fmrt} for the NSE case. It follows from Definition \ref{nsvort} item (3), that if
\be
\label{gammaset}
\Gamma:=\left\{\bu\in V; E_1\leq \left| \bu\right|^2+\alpha^2\left\|\bu\right\|^2 < E_2\right\},
\ee
then, by the Cauchy-Schwarz and Poincar\'e, \eqref{Poinc},  inequalities,
\[
\begin{aligned}
&\nu\int_{\Gamma}\left\| \bu\right\|^2d\mu^\alpha(\bu)\leq\int_{\Gamma}\left|\mathbf{f}\right|\left| \bu\right|d\mu^\alpha(\bu)\\
&\leq\left(\int_{\Gamma}\frac{\left|\mathbf{f}\right|}{\lambda_1^{1/2}}\left\| \bu\right\|^2 d\mu^\alpha(\bu)\right)^{1/2}\\
&\leq\left(\int_{\Gamma}\frac{\left|\mathbf{f}\right|^2}{\lambda_1}\; d\mu^\alpha(\bu)\right)^{1/2}\left(\int_{\Gamma}\left\| \bu\right\|^2\, d\mu^\alpha(\bu)\right)^{1/2},
\end{aligned}
\]
where $\lambda_1$ is the first eigenvalue of the  operator $A$. Thanks to Poncar\'e inequality, \eqref{Poinc}, and
Condition $1$ in Definition \ref{nsvort}, the term in the right-hand side of the last inequality is bounded, and therefore,
\begin{equation}
\label{suppargument}
\int_{\Gamma}\left(\left\| \bu\right\|^2-\frac{\left|\mathbf{f}
\right|^2}{\nu^2\lambda_1}\right)d\mu^\alpha(\bu)\leq 0.
\end{equation}
Notice that if we choose, in the set $\Gamma$, $E_1=0$, and let $E_2\to\infty$, and using the fact that $\mu^\alpha$ is a probability measure, we find that
\be
\la{Aunalpha}
\int_V\left\| \bu\right\|^2d\mu^\alpha(\bu)\leq\frac{\left|\mathbf{f}\right|^2}{\nu^2\lambda_1}.
\ee
Let $0<\epsilon<1$, we define
  $$E_1^\epsilon=\frac{(\lambda_1^{-1}+\alpha^2)\left|\mathbf{f}\right|^2}{\nu^2\lambda_1}(1+\epsilon),$$
   and
    $$E_2^\epsilon=\frac{(\lambda_1^{-1}+\alpha^2)\left|\mathbf{f}\right|^2}{\nu^2\lambda_1}(1+1/\epsilon).$$
     By Poincar\'e inequality, \eqref{Poinc}, we have that $\left| \bu\right|^2+\alpha^2\left\| \bu\right\|^2\leq(\lambda_1^{-1}+\alpha^2)\left\| \bu\right\|^2$, and, therefore,
 \[
 \left\| \bu\right\|^2\geq\frac{\left|\bu\right|^2+\alpha^2\left\|\bu\right\|^2}{\lambda_1^{-1}+\alpha^2}>\frac{\left|\mathbf{f}\right|^2}{\nu^2\lambda_1}(1+\epsilon),
 \]
for all $\bu\in \Gamma^\epsilon:=\left\{\bu\in V;\;E_1^\epsilon\leq|\bu|^2+\alpha^2\left\|\bu\right\|^2<E_2^\epsilon\right\}$. We obtain, by applying \eqref{suppargument} to the set $\Gamma^\epsilon$, that $\mu^{\alpha}(\Gamma^\epsilon)=0$. Letting $\epsilon\downarrow 0$, we conclude \eqref{boundsupp}.$\quad\Box$

\medskip
Now, we define a set that will be often used throughout the work
\be
\label{Bdef}
B_{\alpha}:=\left\{ \bu\in V; \left| \bu\right|^2\leq (\lambda_1^{-1}+\alpha^2)\frac{\left|\mathbf{f}\right|^2}{\nu^2}\right\}.
\ee
The following corollary results immediately from Proposition \ref{boundsupprop}.
\begin{cor}
\label{boundsupp6cor}
Let $\mu^{\alpha}$ be a stationary statistical solution of the Navier-Stokes-Voigt equations, then
\begin{equation}
\label{boundsupp6}
\supp\mu^{\alpha}\subset B_{\alpha}.
\end{equation}
\end{cor}
Next, we show that for every $\alpha>0$, invariant measures for the semigroup $S^{\alpha}(t):V\to V$ generated by the NSV model, \eqref{aueq} (or \eqref{fnseeq}), are stationary statistical solutions for the NSV model. The proof is in the same lines of a related result for the two-dimensional Navier-Stokes presented in \cite{fmrt}.
\begin{prop}
\label{equivalence}
Let $\mu^{\alpha}$ be an invariant measure for the semigroup $S^{\alpha}(t):V\to V$ generated by the NSV model, \eqref{aueq} (or \eqref{fnseeq}). Then,  $\mu^{\alpha}$ is a stationary statistical solution of the NSV model, in the sense of Definition \ref{nsvort}.
\end{prop}
\noindent{\bf{Proof.}}
Let $\mu^{\alpha}$ be an invariant measure for the semigroup $S^{\alpha}(t):V\to V$ generated by the NSV model. Then, by Theorem \ref{testethm}, $\supp\mu^{\alpha}$ is included in the global attractor, $\mathcal{A}$, of $S^{\alpha}$, which is bounded in $D(A)$. Therefore, the function $\bu\to\left\|\bu\right\|^2_{D(A)}$ is bounded on the $\supp\mu^{\alpha}$, which implies that
\be
\label{Asupport}
\int_{V}\left\|\bu\right\|_{D(A)}^{2}d\mu^{\alpha}(\bu)<\infty.
\ee
Therefore, Condition $1$ in Definition \ref{nsvort} is satisfied. Now, we prove
Condition $3$. Let $0\leq E_1<E_2<\infty$, and let  $\Gamma=\left\{\bu\in V; E_1\leq \left| \bu\right|^2+\alpha^2\left\|\bu\right\|^2 < E_2\right\}$. First, we observe that thanks to the Cauchy-Schwarz inequality, the Poincar\'e inequality, \eqref{Poinc}, and to \eqref{Asupport}, the map $\bu\to (\nu\left\|\bu\right\|^2-(\mathbf{f},\bu))$ belongs to $L^1(\mu^\alpha)$. Since $\mu^\alpha$ is an invariant measure, we have
\[
\int_{\Gamma} \left(\nu\left\|\bu\right\|^2-(\mathbf{f},\bu)\right)\, d\mu^{\alpha}(\bu)= \int_{\Gamma}\left( \nu\left\|S^{\alpha}(t)\bu\right\|^2-(\mathbf{f},S^{\alpha}(t)\bu)\right)\, d\mu^{\alpha}(\bu),
\]
for all $t\geq 0$. Now, take the average with respect to $t$ over $[0,T]$, and use the fact that the left-hand side of the above equation is constant in time, to obtain
\be
\begin{aligned}
\int_{\Gamma} \left(\nu\left\|\bu\right\|^2-(\mathbf{f},\bu)\right)\, d\mu^{\alpha}(\bu)&= \frac{1}{T}\int_0^T\int_{\Gamma} \left(\nu\left\|S^{\alpha}(t)\bu\right\|^2-(\mathbf{f},S^{\alpha}(t)\bu)\right)\, d\mu^{\alpha}(\bu)dt\\
&= \int_{\Gamma} \frac{1}{T}\int_0^T\left(\nu\left\|S^{\alpha}(t)\bu\right\|^2-(\mathbf{f},S^{\alpha}(t)\bu)\right)\, dt\,d\mu^{\alpha}(\bu),
\end{aligned}
\ee
where in the last step we are allowed to use the Fubini theorem, (see, e.g., \cite{fmrt}), because the integrand is bounded in $\Gamma$,  jointly continuous on $[0,T]\times V$, and belongs to $L^1([0,T]\times V)$, for all $T>0$. By the energy equality \eqref{enequation}, we obtain
\be
\begin{aligned}
&\int_{\Gamma} \left(\nu\left\|\bu\right\|^2-(\mathbf{f},\bu)\right)\, d\mu^{\alpha}(\bu)\\
&= \frac{1}{2T}\int_{\Gamma} \left(\left|\bu\right|^2+\alpha^2\left\|\bu\right\|^2 -\left|S^{\alpha}(T)\bu\right|^2 -\alpha^2\left\|S^{\alpha}(T)\bu\right\|^2\right)\, d\mu^{\alpha}(\bu),
\end{aligned}
\ee
for all $T>0$. Since $t\to S^{\alpha}(t)\bu$ is bounded in $V$ over $[0,\infty)$, then by letting $T\to \infty$ we obtain by the Lebesgue dominated convergence theorem that
\[
\int_{\Gamma} \left(\nu\left\|\bu\right\|^2-(\mathbf{f},\bu)\right)\, d\mu^{\alpha}(\bu)=0,
\]
which is Condition $3$ in Definition \ref{nsvort}.
For every $m\in\mathbf{Z}^+$, let us define the projector $P_m:H\to H$ by
\[
P_m \bu:=\sum_{j=1}^m(\bu,\mathbf{w}_j)\bw_j, \quad\forall \bu\in H,
\]
where $\left\{\mathbf{w}_j\right\}$ is an orthonormal basis of $H$ consisting of eigenvectors of $A$. We now prove
Condition $2$ of Definition \ref{nsvort}. Let $\Phi\in \mathcal{T}^\alpha$ be a test functional.  For $m\in\mathbb{Z}^+$, we define $\Phi_m(\bu):=\Phi(P_m\bu)$. It is easy to see that $\Phi_m'(\bu)=\Phi'(P_m\bu)P_m$ is a $C^1$ functional on $V$. Because $\mu^\alpha$ is invariant, we have
\be
\begin{aligned}
&\int_V ((I+\alpha^2A)^{-1}\mathbf{F}(\bu),\Phi_m'(\bu)) d\mu^{\alpha}(\bu)\\
&=\int_V ((I+\alpha^2A)^{-1}\mathbf{F}(S^\alpha(t)\bu),\Phi_m'(S^\alpha(t)\bu)) d\mu^{\alpha}(\bu)\\
\end{aligned}
\ee
Now, using the fact that the left-hand side is constant in time, we take the average with respect to $t$ over $[0,T]$, to obtain
\be
\la{lastrolha}
\begin{aligned}
&\int_V ((I+\alpha^2A)^{-1}\mathbf{F}(\bu),\Phi_m'(\bu)) d\mu^{\alpha}(\bu)\\
&=\frac{1}{T}\int_0^T\int_V ((I+\alpha^2A)^{-1}\mathbf{F}(S^\alpha(t)\bu),\Phi_m'(S^\alpha(t)\bu)) d\mu^{\alpha}(\bu)\,dt\\
&=\int_V\frac{1}{T}\int_0^T ((I+\alpha^2A)^{-1}\mathbf{F}(S^\alpha(t)\bu),\Phi_m'(S^\alpha(t)\bu))\,dt\, d\mu^{\alpha}(\bu).
\end{aligned}
\ee
where we have used Fubini's theorem in the last step.

Now, if $\bu\in D(A)$, then by the global regularity result of the NSV proved in \cite{ctk, oskolkov1}, see also \cite{klt}, the function $t\to (I+\alpha^2A)S^{\alpha}(t)\bu$ belongs to $C^1((0,\infty),H)$. Therefore, because the operator $(I+\alpha^2A)^{-1}:H\to H$ is bounded, see Lemma \ref{opbound}, we obtain, by \eqref{fnseeq}, that $S^{\alpha}(t)(\bu)$ satisfies
 \be
 \frac{d}{dt}S^{\alpha}(t)(\bu)=(I+\alpha^2A)^{-1}\mathbf{F}(S^\alpha(t)(\bu))\quad \mbox{in}\; H.
 \ee
 Therefore,
\be
\la{cond2nsv}
\begin{aligned}
&\frac{d}{dt}\Phi_m(S^\alpha(t)(\bu))=(\Phi_m'(S^\alpha(t)(\bu)),\frac{d}{dt}S^\alpha(t)(\bu))\\
&=(\Phi_m'(S^\alpha(t)(\bu)),(I+\alpha^2A)^{-1}\mathbf{F}(S^\alpha(t)(\bu))),
\end{aligned}
\ee
for all $t> 0$, and any $\bu\in D(A)$, where $\mathbf{F}$ is as in \eqref{fnseeq}. Then, take the average with respect to $t$ over $[0,T]$, and substituting the result in the last line of \eqref{lastrolha}, we conclude that
\be
\int_V ((I+\alpha^2A)^{-1}\mathbf{F}(\bu),\Phi_m'(\bu)) d\mu^{\alpha}(\bu)=\int_V \frac{1}{T}[\Phi_m(S^\alpha(T)\bu)-\Phi_m(\bu)]d\mu^{\alpha}(\bu).
\ee
Because $\Phi_m$ is bounded, we obtain, as $T\to \infty$,
\be
\int_V ((I+\alpha^2A)^{-1}\mathbf{F}(\bu),\Phi_m'(\bu)) d\mu^{\alpha}(\bu)=0.
\ee
Since $\Phi_m'(\bu)=\Phi'_m(P_m\bu)P_m$, we have that
\be
\int_V (P_m(I+\alpha^2A)^{-1}\mathbf{F}(\bu),\Phi'(P_m\bu)) d\mu^{\alpha}(\bu)=0.
\ee
Now, because $P_m\bu\to \bu$, in $H$, as $m\to \infty$, for all $\bu\in V$, we have
\be
(P_m(I+\alpha^2A)^{-1}\mathbf{F}(\bu),\Phi'(P_m\bu)) \to ((I+\alpha^2A)^{-1}\mathbf{F}(\bu),\Phi'(\bu)), \textit{ as }m\to\infty.
\ee
Therefore, because $\Phi'$ is bounded, we have by \eqref{b5} that there exists a  constant $c>0$, which is independent of $\alpha$, such that
\[
\left|(P_m(I+\alpha^2A)^{-1}\mathbf{F}(\bu),\Phi'(P_m\bu))\right|< c(\left\|\bu\right\|^2_{D(A)}+|\mathbf{f}|),
\]
where the right-hand side belongs to  $L^1(d\mu^\alpha)$, by \eqref{Asupport}. Then, by the Lebesgue dominated convergence theorem, we obtain
\be
\begin{aligned}
&\int_V((I+\alpha^2A)^{-1}\mathbf{F}(\bu),\Phi'(\bu))d\mu^{\alpha}(\bu)=\\
&\lim_{m\to\infty}\int_V(P_m(I+\alpha^2A)^{-1}\mathbf{F}(\bu),\Phi'(P_m\bu))d\mu^{\alpha}(\bu)=0.
\end{aligned}
\ee
Therefore, Condition $2$ holds, completing the proof of the theorem.$\quad\Box$

\begin{cor}
\la{boundsupp7}
The set of stationary statistical solutions of the NSV equations is nonempty.
\end{cor}
\noindent{\bf{Proof.}} As we have observed earlier, the set of steady state solutions of the NSV, or equivalently of the NSE, is non-empty, see, e.g., \cite{constantinfoias, temam}. Therefore, the Dirac delta measures concentrated on the steady state solutions and all their convex combinations are invariant measures. Consequently, by Proposition \ref{equivalence}, they are stationary statistical solutions of the NSV model.

\medskip

\noindent{\bf{Remark.}} One could also prove the converse of Proposition \ref{equivalence}, i.e.,
that every stationary statistical solution of the NSV model is an invariant measure, by following the arguments of Foias in \cite{foias72, foias73} for the $2$D periodic case, see also \cite{fmrt}. Only a slight modification in the treatment of the nonlinear term is necessary, but which can be circumvented by using that $\supp\mu^\alpha$ is included in $D(A)$, and is bounded in $V$. We do not present it here, since the semigroup generated by the NSV model is well defined in $V$, and therefore, we can use the more natural notion of invariant measures in our approximation results in Section $6$.

In the sequel, we use the symbol $\langle\ldots\rangle^\alpha$ to denote average with respect to an invariant measure of the  NSV model $\mu^\alpha$, i.e.
 \[
 \langle\varphi\rangle^\alpha:=\int_V\varphi(\bu)d\mu^\alpha(\bu);\quad\forall \varphi\in L^1(d\mu^\alpha).
 \]

 We can use the stationary statistical solutions formalism to show that the stationary Reynolds averaged equations hold for the NSV equations in this framework. Indeed,  we can define the mean velocity field $\langle\bu\rangle^\alpha\in D(A)$, thanks to
 Condition $1$ in Definition \eqref{nsvort}, as follows
\[
 \left(\langle\bu\rangle^\alpha,\bv\right)=\int_{V}(\bu,\bv)d\mu^\alpha(\bu)=\int_{D(A)}(\bu,\bv)d\mu^\alpha(\bu),\quad\forall \bv\in D(A)'.
\]
Moreover, thanks to \eqref{b6} and Condition $1$ of Definition
\ref{3sssvort}, we may define the average $\langle
B(\bu,\bu)\rangle^\alpha\in H$ by
\[
 (\langle B(\bu,\bu)\rangle^\alpha,\bv)=\int_{V}(B(\bu,\bu),\bv)d\mu^\alpha(\bu)=\int_{D(A)}(B(\bu,\bu),\bv)d\mu^\alpha(\bu), \,
       \forall \bv\in H.
\]

The mean flow $\langle\bu\rangle^\alpha$ is a vector field on $\Omega$ with
$\langle\bu\rangle^\alpha\in D(A)$, while $\langle B(\bu,\bu)\rangle^\alpha\in H$.
The next proposition shows that, assuming statistical equilibrium, the  vector field $\langle\bu\rangle^\alpha$ also satisfies the Reynolds averaged equations in $H$. This should be compared to the NSE case, where these equations are known to be valid only in $V'$, (see, e.g., \cite{fmrt}).

\begin{prop}
Let $\mu^\alpha$ be an invariant measure of the NSV equations. Then, the following functional form of the Reynolds averaged equations holds in $H$:
\begin{equation}
\label{Reynoldseq}
\nu A\langle\bu\rangle^{\alpha}+\langle B(\bu,\bu)\rangle^{\alpha}=\mathbf{f}.
\end{equation}
\end{prop}

\noindent{\bf{Proof.}} We will use here the notation of Section $3$, and set $\kappa_1^2=\lambda_1$. Let $\psi$ be a $C^{1}$ real-valued function with compact
support on $\mathbb{R}$. For any $\bv\in H$, any finite wavenumber
$\kappa\geq \kappa_1$, and every $\bu \in D(A)$, the real-valued function
\be
\label{psirey}
\Phi(\bu)=\psi(((I+\alpha^2A)^{1/2}\bu,(I+\alpha^2A)^{1/2}\bv_{\kappa_{1},\kappa}))
\ee
 is a cylindrical test function, and it satisfies
\[
(\Phi'(\bu),\phi)=\psi'(((I+\alpha^2A)^{1/2}\bu,(I+\alpha^2A)^{1/2}\bv_{\kappa_{1},\kappa})) ((I+\alpha^2A)\phi,\bv_{\kappa_{1},\kappa}).\\
\]
for every $\phi\in D(A)$. Thus, considering $\phi=(I+\alpha^2A)^{-1}F(\bu)$, which belongs to $D(A)$, we obtain
\[
(\Phi'(\bu),(I+\alpha^2A)^{-1}F(\bu))=\psi'(((I+\alpha^2A)^{1/2}\bu,(I+\alpha^2A)^{1/2}\bv_{\kappa_{1},\kappa})) (F(\bu),\bv_{\kappa_{1},\kappa}).
\]
By Proposition \ref{equivalence}, $\mu^{\alpha}$ is also a stationary statistical solutions of the NSV model, therefore, by substituting the expression above in
Condition $2$ of Definition \ref{nsvort}, we obtain
\[
\begin{aligned}
\int_{V}\psi'(((I+\alpha^2A)^{1/2}\bu,&(I+\alpha^2A)^{1/2}\bv_{\kappa_{1},\kappa}))\left\{(\mathbf{f},\bv_{\kappa_{1},\kappa})\right.\\
&\left.-\nu(A\bu,\bv_{\kappa_{1},\kappa})-b(\bu,\bu,\bv_{\kappa_{1},\kappa})\right\}d\mu^\alpha(\bu)=0.
\end{aligned}
\]
Now, by virtue of  Proposition \ref{boundsupprop}, $\supp\mu^\alpha$
is bounded in $V$,  and therefore the map $\bu\in \supp\mu^\alpha
\to
((I+\alpha^2A)^{1/2}\bu,(I+\alpha^2A)^{1/2}\bv_{\kappa_{1},\kappa})$
is uniformly bounded in $\supp\mu^\alpha$, let us say by $M>0$.
Therefore, we can choose $\psi$, above, such that $\psi'\equiv 1$ at
the interval $[-M, M]$, therefore,
$\psi'(((I+\alpha^2A)^{1/2}\bu,(I+\alpha^2A)^{1/2}\bv_{\kappa_{1},\kappa}))=1$,
for every $\bu\in \supp\mu^\alpha$. This yields
\[
\int_{V}\left\{(\mathbf{f},\bv_{\kappa_{1},\kappa})-\nu(A\bu,\bv_{\kappa_{1},\kappa})-b(\bu,\bu,\bv_{\kappa_{1},\kappa})\right\}d\mu^\alpha(\bu)=0.
\]
For each fixed $\bv\in H$, we observe that since $\supp\mu^\alpha\subset D(A)$, we have by the Cauchy-Schwarz inequality, and by \eqref{b5}, that
\[
|(\mathbf{f},\bv_{\kappa_{1},\kappa})-\nu(A\bu,\bv_{\kappa_{1},\kappa})-b(\bu,\bu,\bv_{\kappa_{1},\kappa})|\leq |\mathbf{f}||\bv|+\nu|A\bu||\bv|+C|A\bu|^2|\bv|.
\]

Thanks to Condition $1$ of Definition \ref{nsvort}, we can apply the Lebesgue dominated convergence Theorem, and let $k\to\infty$ to obtain
\[
\int_{V}\left\{(\mathbf{f},\bv)-\nu(A\bu,\bv)-b(\bu,\bu,\bv)\right\}d\mu^\alpha(\bu)=0,
\]
which yields the result. $\quad\Box$

\section{Averaged Energy Budget for the NSV model}

In this section, we follow \cite{fmrt} to investigate the energy distribution scale-by-scale for the $3$D NSV equations. For the Navier-Stokes equations, assuming that there exists an inertial range, i.e. an extensive range of wavenumbers where the viscous dissipation does not play a significant role, one can show that the energy simply cascades through these wavenumbers with a rate equal to  the mean energy dissipation rate, $\epsilon=\nu\langle\left\|\bu\right\|^2\rangle^\alpha$. For the NSV equations, a similar scenario holds for the $\alpha$-energy
\[
\mathcal{E}_{\alpha}=\frac{1}{2}\left|\bu\right|^2+\frac{\alpha^2}{2}\left\|\bu\right\|^2.
\]

As it is usual in the studies of homogeneous turbulence, and using the notation of Section $3$, we will consider the forcing $\mathbf{f}$ with finite eigenmodes, i.e.
\be
\mathbf{f}=\sum_{\underline{\kappa}^2\leq \lambda_j\leq\bar{\kappa}^2}(\mathbf{f},\mathbf{w}_j)\mathbf{w}_j,
\ee
for the orthonormal basis, $\left\{\mathbf{w}_j\right\}$, of $H$ consisting of the eigenvectors of $A$, with $\underline{\kappa}\geq\kappa_1=\sqrt{\lambda_1}$.

Now, we argue as in the proof of Proposition \ref{equivalence} to obtain an averaged energy balance equation. We skip the details, but we sketch the derivation: Using the fact that $\mu^\alpha$ is an invariant measure, using equation \eqref{projected},  averaging with respect to $t$ over $[0,T]$, and using Fubini's Theorem, we have
\be
\begin{aligned}
&\int_{\Gamma} \left(\nu\left\|\bu_{\kappa',\kappa''}\right\|^2-(\mathbf{f},\bu_{\kappa',\kappa''})-[e^{\alpha}_{\kappa'}(\bu)-e^{\alpha}_{\kappa''}(\bu)]\right) d\mu^{\alpha}(\bu)\\
&= \frac{1}{2T}\int_{\Gamma}
\left(\left|\bu_{\kappa',\kappa''}\right|^2+\alpha^2\left\|\bu_{\kappa',\kappa''}\right\|^2
-\left|S^{\alpha}(t)\bu_{\kappa',\kappa''}\right|^2
-\alpha^2\left\|S^{\alpha}(t)\bu_{\kappa',\kappa''}\right\|^2\right)
d\mu^{\alpha}(\bu)\,,
\end{aligned}
\ee where $\Gamma$ is as in the proof of Proposition
\ref{equivalence}.
Now, let
$\kappa''\geq \kappa' >\bar{\kappa}$, and letting $T\to \infty$ in
the above equation, we obtain the following balance equation
\be
\nu\langle\left\|\bu_{\kappa',\kappa''}\right\|^2\rangle^\alpha=\langle e^{\alpha}_{\kappa'}(\bu)-e^{\alpha}_{\kappa''}(\bu)\rangle^{\alpha}.
\la{tbud5}
\ee

The expression on the right-hand side of the last equality is the mean net $\alpha$-energy transfer in the energy shell $[\kappa',\kappa'']$. Because the $\supp\mu^{\alpha}$ in included in $D(A)$ and is bounded in $V$, we can use the Lebesgue dominated convergence Theorem, and let $\kappa''\to\infty$ to obtain
\be
\nu\langle\left\|\bu_{\kappa',\infty}\right\|^2\rangle^\alpha=\langle e_{\kappa'}(\bu)\rangle^{\alpha}.
\la{infitransf}
\ee
This expression shows that the net $\alpha$-energy transfer is positive for every $\kappa>\bar{\kappa}$.
We remark that a similar expression holds for the three-dimensional NSE case substituting equality by inequality, see, e.g., \cite{fmrt}.  Therefore, we have proved
\begin{prop}
\la{entransf}
Let $\mu^\alpha$ be an invariant measure for the semigroup, $\left\{S^\alpha(t)\right\}_{t\geq 0}$, generated by the NSV model, and let $\bar{\kappa}<\kappa'\leq\kappa''$. Then
\be
\la{inertcond3}
\nu\langle\left\|\bu_{\kappa',\kappa''}\right\|^2\rangle^{\alpha}=(\langle e_{\kappa'}(\bu)\rangle^{\alpha}-\langle e_{\kappa''}(\bu)\rangle^{\alpha}).
\ee
Moreover, for all $\kappa>\bar{\kappa}$, we have
\be
\nu\langle\left\|\bu_{\kappa,\infty}\right\|^2\rangle^\alpha=\langle e_{\kappa}(\bu)\rangle^\alpha.
\ee
\end{prop}
The result above shows that the mean net $\alpha$-energy transfer is positive, which is an important consistency check for turbulence models. Moreover, if we assume that there exists a range of wavenumbers, $\left[\kappa',\kappa''\right]$, where the viscous dissipation term satisfies $\nu\langle\left\|\bu_{\kappa',\kappa''}\right\|^2\rangle^\alpha\ll 1$, it follows from \eqref{inertcond3} that the $\alpha$-energy transfer is nearly constant within this range, that is, there exists an inertial range for the NSV model.

\section{Statistical approximations of the $3$D Navier-Stokes equations}

In this section we prove that given a sequence of invariant measures for the NSV equations, $\left\{\mu^{\alpha_n}\right\}$, with the regularizing parameter $\alpha_n$ converging to $0$, as $n\to\infty$, there exists a subsequence converging weakly to a \textit{strong stationary statistical solutions} of the $3$D Navier-Stokes equations, a notion that we introduce in Definition $3$ below.  This is a slight modification of the notion of stationary statistical solutions of the $3$D NSE introduced by Foias in \cite{foias72, foias73}, see also \cite{fmrt}.

The main difficulty in proving this result is that the supports of the invariant measures, $\mu^\alpha$, of the NSV are not uniformly bounded, with respect to the parameter $\alpha$, in the norm of $V$, as $\alpha$ tends to zero. This means that we cannot use in a straightforward manner the standard Prokhorov's Theorem, see, e.g.  \cite[Thm. 1.12]{prokhorov}. This is because the whole space $V$ is not metrizable when endowed with its weak topology. Therefore, we have to consider more general theorems concerning convergence of measures in arbitrary Hausdorff spaces. Fortunately, the family of measures considered in this section satisfy the tightness condition of \cite{topsoe}, and the weak convergence holds for our case.

We start by defining the terminology, from \cite{topsoe}, that is relevant to the results presented in this section. A paving is a non-empty set consisting of subsets of a given set $X$. $\mathcal{N}\subset 2^{X}$ is said to be a $(\emptyset,\cup f, \cap c)-$paving of $X$ if $\mathcal{N}$ is closed under finite unions and countable intersections, and if in addition $\emptyset\in \mathcal{N}$. Similarly, $\mathcal{N}\subset 2^{X}$ is said to be a $(\emptyset, \cup f, \cap f)-$paving of $X$ if $\mathcal{N}$ is closed under finite unions and finite intersections, and $\emptyset\in \mathcal{N}$. A paving $\mathcal{K}$ is called compact [semicompact] if every  family [every countable family] of sets in $\mathcal{K}$, which has the finite intersection property (the intersection of any finite subset of $\mathcal{K}$ is non-empty) has a non-empty intersection (the intersection of all elements of $\mathcal{K}$ is non-empty). A paving $\mathcal{L}$ separates the sets in $\mathcal{K}$ if to any pair $K_1$, $K_2$ of disjoint elements in $\mathcal{K}$, we can find a pair $G_1$, $G_2$ of disjoint sets in $\mathcal{L}$ such that  $K_1\subset G_1$ and $K_2\subset G_2$.

The results stated in \cite{topsoe} consider a Hausdorff topological space $X$, and a net, $\left\{x_\beta\right\}$, in X. The notion of net can be found in \cite{topsoe}, but it is not important for our present work because the space $V$ is a separable topological space with its weak topology, and thus all statements can be done in terms of sequences. However,  we will state a theorem appearing in \cite{topsoe} in its original formulation, in terms of nets. We only recall that a net $(x_\beta)_{\beta\in D}$ on a topological space $X$ is compact if every subnet has a further subnet that converges.

  Let $X$ be a Hausdorff space, and let $\mathcal{L}$ and $\mathcal{K}$ be  pavings in $X$. We denote by $\mathcal{B}=\mathcal{B}(X,\mathcal{L})$, the smallest $\sigma$-field containing every set $E\subset X$ for which $L\cap E\in \mathcal{L}$, $\forall L\in \mathcal{L}$. We denote by $\mathcal{M}_+(X,\mathcal{L})$ the set of finite, non-negative measures defined on $\mathcal{B}(X,\mathcal{L})$. $\mathcal{M}_+(X,\mathcal{L},\mathcal{K})$ denotes the set of measures in $\mathcal{M}_+(X,\mathcal{L})$ which are regular with respect to $\mathcal{K}$, i.e., $\mu\in \mathcal{M}_+(X,\mathcal{L},\mathcal{K})$ if and only if $\mu$ satisfies
\be
\la{regdef}
\mu(E)=\sup\left\{\mu(K);\; K\in\mathcal{K},\, K\subset E \right\},\quad\forall E\subset \mathcal{B}.
\ee
From now on, we will consider the following axioms
 \begin{enumerate}
\item
$X$ is a Hausdorff topological space.
\item
\la{top1}
$\mathcal{K}$ is a $(\emptyset, \cup f, \cap c)-$paving of $X$.
\item
\la{top2}
$\mathcal{L}$ is a $(\emptyset, \cup f, \cap f)-$paving of $X$.
\item
\la{top3}
$K\setminus G\in\mathcal{K}$, $\forall K\in\mathcal{K}$, $\forall G\in \mathcal{L}$.
\item
\la{top4}
$\mathcal{L}$ separates the sets in $\mathcal{K}$.
\item
\la{top5}
$\mathcal{K}$ is semicompact.
\end{enumerate}

Let $\mu\in \mathcal{M}_+(X,\mathcal{L})$, and let $\left\{\mu^\beta\right\}$ be a net in $\mathcal{M}_+(X,\mathcal{L})$. Then, we say that $\left\{\mu^\beta\right\}$ converges to $\mu$ in the weak topology if and only if $\mu^{\beta}(X)\to \mu(X)$ and $\liminf\mu^\beta(G)\geq\mu(G)$, $\forall G\in \mathcal{L}$.

Now, we are going to state the abstract result that we need to continue with our investigation. It is just the first half of Corollary $1$ in \cite{topsoe}.

\begin{thm}(Corollary 1, \cite{topsoe})
\la{absconv}
Assume that axioms $I-VI$ are satisfied and let $(\mu^\beta)_{\beta\in D}$ be a net on $\mathcal{M}_+(X,\mathcal{L},\mathcal{K})$ such that $\limsup_{\beta\in D} \mu^\beta(X)< \infty$.
If the tightness condition
\be
\la{tightcond}
\inf_{K\in\mathcal{K}}\limsup_{\beta\in D}\mu^\beta(X\setminus K)=0
\ee
holds, then $(\mu^\beta)_{\beta\in D}$ is compact.
\end{thm}

We will now translate all the preceding abstract definitions to our specific context. We consider the topological space $X=V$, where $V$ is defined in \eqref{vdefin}, endowed with its weak topology. Thanks to the Hahn-Banach Theorem, the space $V$ is a Hausdorff topological space with the weak topology. Moreover, $V$ is also separable with the weak topology, see, e.g., \cite{fmrt}. This implies that there is no need to use nets in the application of Theorem \ref{absconv} in this work, and we will consider only sequences in the subsequent results.

We denote by $\mathcal{K}(V)$, $\mathcal{F}(V)$, $\mathcal{L}(V)$ and $\mathcal{B}(V)$ the pavings of $V$ of weakly compact sets, weakly closed sets, weakly open sets, and weakly Borel sets, respectively. By $\mathcal{M}_+(V,\mathcal{L}(V))$  we denote the space of finite non-negative measures defined on $\mathcal{B}=\mathcal{B}(V,\mathcal{L}(V))$. It is easy to prove that the axioms $I-VI$ are valid for the above choice of $\mathcal{K}(V)$, $\mathcal{F}(V)$, $\mathcal{L}(V)$ and $\mathcal{B}(V)$. We omit the proof of these facts.

 The weak topology on $\mathcal{M}_+(V,\mathcal{B}(V))$  is defined via the paving $\mathcal{L}(V)$, and it is exactly the familiar weak topology in the well-known Polish spaces, i.e. $\mu^\beta$ converges in the weak topology to $\mu$ if and only if
 $$
 \int_V h(\bu) d\mu^\beta(\bu)\to  \int_V h(\bu) d\mu(\bu),
 $$
  for every $h:V\to \mathbb{R}$ bounded continuous function. We also recall  that the subset of Borel probability measures $\mathcal{M}_1$ is closed in $\mathcal{M}_+(V)$ under the weak topology, (see, e.g., \cite{DS}), which implies that weakly converging sequences in $\mathcal{M}_1$ converge to elements in $\mathcal{M}_1$.

Now we will prove that every Borel finite measure defined on $V$ endowed with the weak topology satisfies the tightness condition of Theorem \ref{absconv}.

\begin{lemma}
\la{boreltight}
Let $\mu\in  \mathcal{M}_+(V,\mathcal{L}(V))$, then  $\mu\in\mathcal{M}_+(V,\mathcal{L}(V),\mathcal{K}(V))$.
\end{lemma}
\noindent{\bf{Proof.}}  First, we notice that because $V$ with the strong topology is a separable Hilbert space, the Borel $\sigma$-algebra associated with the strong topology is the same as the Borel $\sigma$-algebra associated with the weak topology, (see, e.g., \cite{fmrt}).

Moreover, because $V$ with the strong topology is separable, any strongly Borel probability measure, $\mu$, is regular in the sense that for every strongly Borel set $E$,
\be
\la{regdef2}
\mu(E)=\sup\left\{\mu(K);\; K \mbox{ is strongly compact },\, K\subset E \right\},
\ee
see, e.g., \cite{DS}. Now, because strongly compact sets in $V$ are also weakly compact, we have that if $E$ is a weakly Borel set in $V$ (which also implies that it is a strongly Borel set), then $\mu$ satisfies
\be
\la{regdef3}
\begin{aligned}
\mu(E)&\geq\sup\left\{\mu(K);\; K \mbox{ is weakly compact },\, K\subset E \right\}\\
&\geq\sup\left\{\mu(K);\; K \mbox{ is strongly compact },\, K\subset E \right\}=\mu(E).
\end{aligned}
\ee
Therefore,
\be
\mu(E)=\sup\left\{\mu(K);\; K \mbox{ is weakly compact },\, K\subset E \right\},
\ee
which shows that $\mu\in\mathcal{M}_+(V,\mathcal{L}(V),\mathcal{K}(V))$. $\quad\Box$

\medskip

Now, it is easy to see that Theorem \ref{absconv},  Lemma \ref{boreltight}, and the preceding discussions imply the following theorem:

\begin{thm}
\label{absbor}
Let $\mathcal{K}$ denote the paving of $V$ formed by the weakly compact sets.  Consider the space $V$ endowed with its weak topology, and let $\left\{\mu_n\right\}$ be a sequence of Borel probability measures in $V$.
If the tightness condition
\be
\la{Vtight}
\inf_{K\in\mathcal{K}}\limsup_{n\to\infty}\mu_n(V\setminus K)=0
\ee
holds, then there exists a subsequence, also denoted by $\left\{\mu_n\right\}$, such that $\mu_n\rightharpoonup\mu$ weakly  in $\mathcal{M}_+(V,\mathcal{L}(V))$.
\end{thm}

Because of the convergence result of our main theorem in this section occurs in the space $V$, endowed with its weak topology, the limit measure thus obtained will be defined on this space as well. Therefore, we need to define the following notion

\begin{definition}
\label{3sssvort}

A strong stationary statistical solution of the $3$D Navier-Stokes equations is a Borel probability measure $\mu$ on $V$ such that
\begin{enumerate}

\item[(1)]
\label{3dssv1}
$\displaystyle{\int_{V}\left\|\bu\right\|^{2}d\mu(\bu)<\infty;}$

\item[(2)]
\label{3dssv2}
$\displaystyle{\int_{V} \langle\Psi'(\bu),\mathbf{f}-\nu A\bu -B(\bu,\bu)\rangle_{V, V'}\, d\mu(\bu)=0,}
\\
      \text{for any test functional}\; \Psi\in\mathcal{T}, \text{where $\mathcal{T}$ is defined in Definition } \ref{3test}.$

\item[(3)]
\label{3dssv3}
$
\displaystyle{\int_{ E_1\leq \left| \bu\right|^2< E_2}\left[\nu\left\|\bu\right\|^2-( \bbf,\bu)\right]d\mu(\bu)\leq 0}, \, \mbox{for every}\; 0 \leq E_1 < E_2<\infty.$

\end{enumerate}
\end{definition}
As remarked before, we may use the weakly Borel $\sigma$-algebra in Definition \ref{3sssvort} because $V$ is a separable Hilbert space, and therefore, weakly Borel sets coincide with the strong Borel sets, see, e.g., \cite{DS}.

\begin{definition}
\label{3test}
The class of test functions $\mathcal{T}$ is the set of functions $\Psi:V\rightarrow\mathbb{R}$ of the form
\be
\label{3Itest}
\Psi(\bu):=\psi\left(( \bu,\mathbf{g}_{1}),\ldots,( \bu,\mathbf{g}_{m})\right),
\ee
where the function $\psi$ is a $C^{1}$ scalar valued function defined on $\mathbb{R}^{m}$, with $m\in\mathbb{Z}^{+}$, and
$\mathbf{g}_{1},\ldots,\mathbf{g}_{m}$ belong to $V$.
\end{definition}

\medskip

\noindent{\bf{Remark\,1:}} We remark again that for $\mathbf{f}\in
H$, there exist steady state solutions of the $3$D Navier-Stokes
equations which belongs to $D(A)$, see, e.g., \cite{constantinfoias,
temam}. Therefore, every Dirac measure concentrated on a steady
solution of the $3$D Navier-Stokes equations which belongs to D(A)
is a strong stationary statistical solution. Moreover, if $\bu(t)$
is a strong solution of the NSE, bounded in $V$ for every $t\geq 0$,
then the trajectory of $\bu(t)$ is included in a closed ball in $V$,
which is compact in the weak topology of $V$. Therefore, one can
also generate a strong stationary statistical solution by the
Krylov-Bogolyubov procedure for generating invariant measures; see,
e.g., \cite{cr}, \cite{fmrt} for details on how to use this
procedure to generate time-average measures and the Banach limit.

\noindent{\bf{Remark \,2:}} We remark that strong stationary
statistical solutions are defined over the space $V$ with its
corresponding Borel $\sigma$-algebra, while that the usual
stationary statistical solutions, defined by Foias  in
\cite{foias72}, \cite{foias73}, are defined over the space $H$ with
its Borel $\sigma$-algebra.   This justifies the denomination strong
stationary statistical solutions. Indeed, because Leray-Hopf weak
solutions of the $3$D NSE  are uniformly bounded in the $H$-norm,
with respect to time, one can generate, by the Krylov-Bogolyubov
procedure, a Borel probability measure defined over $H$, which is a
stationary statistical solution, (see, e.g., \cite{fmrt}). However,
because strong stationary statistical solutions are defined over
$V$, the Krylov-Bogolyubov procedure can generate a strong
stationary statistical solution only if applied to strong solutions
of the $3D$ NSE, that are uniformly bounded, with respect to time,
in the $V$-norm, as described in the former remark.

\medskip
\medskip
We now proceed with the computation of $\Psi'$ for test functions $\Psi\in\mathcal{T}$. It is easy to see that for $\Psi\in\mathcal{T}$, as defined in \eqref{3Itest}, we have
\[
\nabla_\bu \Psi(\bu)\cdot\phi=\sum_{j=1}^m\partial_j\psi\left((\bu,\mathbf{g}_{1}),\ldots,(\bu,\mathbf{g}_{m})\right) ( \phi,\mathbf{g}_{j}).\\
\]
It is easy to see that $\phi\to\nabla_\bu\Psi(\bu)\cdot\phi$ is a bounded linear continuous functional in  $H$, hence, by the Riesz representation theorem, there exists an element $\Psi'(\bu)\in H$ such that
\[
\nabla_\bu\Psi(\bu)\cdot\phi=(\Psi'(\bu),\phi),\quad\forall\phi\in H.
\]
This is the identification that is used in Definition \ref{3sssvort}. More precisely,
\be
\label{ftest1}
\Psi'(\bu)=\sum_{j=1}^m\partial_j\psi\left((\bu,\mathbf{g}_{1}),\ldots,(\bu,\mathbf{g}_{m})\right)\mathbf{g}_{j}.
\ee

We now establish a lemma that will play a key role in the rest of this work.

\begin{lemma}
\la{opbound}
For every $\alpha\geq 0$, the operators $(I+\alpha^2A)^{-1}:H\to H$ satisfy
\be
\label{unibound}
\left\|(I+\alpha^2A)^{-1}\right\|_{H\to H}\leq 1.
\ee
Moreover, there exists a constant $C>0$, independent of $\alpha$, such that
\be
\la{opconv}
\left|((I+\alpha^2A)^{-1}-I)\phi\right|\leq C\alpha^2 \lambda_1^{1-\eta} |A^{\eta}\phi|,
\ee
for every $\phi\in  D(A^\eta)$, and $\eta>7/4$.

\end{lemma}

\noindent{\bf{Proof.}}
Indeed, let $\left\{\mathbf{w}_j\right\}$ be a complete orthonormal basis of $H$ formed by the eigenvectors of the operator $A$, and let $0<\lambda_1\leq\lambda_2\ldots\leq\lambda_{j}\to\infty$, as $j\to\infty$, be the corresponding eigenvalues of the operator $A$ associated with the eigenvectors $\left\{\mathbf{w}_j\right\}$. Then, by the Parseval identity, for every $\alpha\geq 0$,
\[
|((I+\alpha^2A)^{-1})\phi|^2=\sum_{j=1}^\infty\left|\frac{1}{1+\alpha^2\lambda_j}\right|\left|(\phi,\mathbf{w}_j)\right|^2\leq |\phi|^2,
\]
which proves that $\left\|(I+\alpha^2A)^{-1}\right\|_{H\to H}\leq 1$.
Now, for every $\phi\in H$, we have by the Parseval identity that
\[
\begin{aligned}
&((I+\alpha^2A)^{-1}-I)\phi=\sum_{j=1}^\infty\left(\frac{1}{1+\alpha^2\lambda_j}-1\right)(\phi,\mathbf{w}_j)\mathbf{w}_j\\
&=\sum_{j=1}^\infty\left(-\frac{\alpha^2\lambda_j}{1+\alpha^2\lambda_j}\right)(\phi,\mathbf{w}_j)\mathbf{w}_j.
\end{aligned}
\]
Therefore, because
$|(\phi,\mathbf{w}_j)|^2=\lambda_j^{-2\eta}|(A^\eta\phi,\mathbf{w}_j)|^2\leq
\lambda_j^{-2\eta}|A^\eta\phi|^2$, for all $j\geq 1$, we once again
have by the Parseval identity that \be
\begin{aligned}
\la{Aineq}
&\left|((I+\alpha^2A)^{-1}-I)\phi\right|^2=\sum_{j=1}^\infty\left|\frac{\alpha^2\lambda_j}{1+\alpha^2\lambda_j}\right|^2|(\phi,\mathbf{w}_j)|^2 \\
&=\sum_{j=1}^\infty\left|\frac{\alpha^2\lambda_j^{1-\eta}}{1+\alpha^2\lambda_j}\right|^2|A^{\eta}\phi|^2\leq \alpha^4 \frac{|A^{\eta}\phi|^2}{\lambda_1^{2(\eta-1)}} \sum_{j=1}^\infty\frac{\lambda_1^{2(\eta-1)}}{\lambda_j^{2(\eta-1)}}.\\
\end{aligned}
\ee
The asymptotic behavior of the eigenvalues of the Stokes operator defined in $\Omega \subset \mathbb{R}^n$ is known to satisfy the Weyl-type formula; see, e.g., \cite{babenko}, \cite{metivier}:
\be
\label{Aasympt}
\lambda_k \sim \left(\frac{(2\pi)^n}{\omega_n(n-1)|\Omega|}\right)^{2/n}k^{2/n}, \quad \mbox{ as } k\to\infty,
\ee
where $|\Omega|$ is the n-dimensional Lebesgue measure of $\Omega$, and $\omega_n$ is the volume of the unit ball in $\mathbb{R}^n$, which implies that the infinite series $\sum_{j=1}^\infty\left(\frac{\lambda_1}{\lambda_j}\right)^{2(\eta-1)}$ converges for $\eta>7/4$, if $n=3$.  Therefore, by \eqref{Aasympt}, it is easy to see that
\[
\left|((I+\alpha^2A)^{-1}-I)\phi\right|\leq C \lambda_1^{2(1-\eta)}\alpha^2 |A^{\eta}\phi|,
\]
where $C^2=\sum_{j=1}^\infty\left(\frac{\lambda_1}{\lambda_j}\right)^{2(\eta-1)}$.  $\quad  \Box$

The following lemma, which is a trivial consequence of the compact embedding $V\subset\subset H$, will play a key role in the sequel of this work. We prove it here for the sake of completeness.

\begin{lemma}
\la{hinv}
The function $\bu\to|\bu|$ is weakly continuous with respect to the $V$-topology, and bounded over $B_\alpha\subset V$, $\alpha\in \mathbb{R}^+$, where $B_\alpha$ is as defined in \eqref{Bdef}.
\end{lemma}
\noindent{\bf{Proof.}}  The bound follows immediatelly from the definition of $B_\alpha$ in \eqref{Bdef}. Now, we prove that $\bu\to|\bu|$ is weakly continuous with the $V$-topology. Indeed, suppose that $\bu_j\rightharpoonup \bu$ weakly in $V$, as $j\to\infty$, but $\bu_j$ does not converge strongly to $\bu$ in $H$. Then, there exists $\epsilon>0$, and a subsequence $\left\{\bu_{j_k}\right\}$ of $\left\{\bu_{j}\right\}$ satisfying
\be
|\bu_{j_k}-\bu|>\epsilon,
\la{ws}
\ee
for all $k\in\mathbb{Z}^+$. Now, because  $\bu_{j_k}\rightharpoonup \bu$ weakly in $V$, as $k\to\infty$, we have, by the compact embedding $V\subset\subset H$, that there exists a further subsequence $\left\{\bu_{j_{k_\ell}}\right\}$ of $\left\{\bu_{j_k}\right\}$ converging strongly (of course, also weakly) in $H$ to a vector field $\bv\in V$, as $\ell\to \infty$. Because weak convergence in $V$ implies weak convergence in $H$, $\bu_{j_{k_\ell}}\rightharpoonup \bu$ weakly in $H$, as $\ell \to \infty$. By the uniqueness of the weak limit, $\bu=\bv$. Therefore, $|\bu_{j_{k_\ell}}-\bu|\to 0$, as $\ell\to\infty$, which contradicts \eqref{ws}. This proves the lemma. $\quad\Box$

\medskip

In order to prove the convergence results in the sequel of this work, we need first to consider a subclass of smoother test functions, $\mathcal{T}_1\subset \mathcal{T}$, consisting of functionals satisfying \eqref{3Itest}, so that the vector fields $\mathbf{g}_i$ appearing in their definitions possess a higher regularity, namely, $\mathbf{g}_i\in \mathcal{V}$, where $\mathcal{V}$ is given in \eqref{smoothf} or \eqref{persmoothf}.

\smallskip

\begin{lemma}
\label{lemma2}
Let $\Psi\in\mathcal{T}_1$. Let $\alpha_0>0$ be fixed, and let $B_{\alpha_0}$ be as defined in \eqref{Bdef}. Then $\Psi'(\bu)\in \bigcap_{\eta\in\mathbb{Z}^+}D(A^\eta)$, and there exists a constant, $C>0$, depending only on $\Psi$, $\ell$, $\eta$  and $\alpha_0$, such that
\be
\label{3testboundt}
|A^\eta\frac{\partial^{\ell}}{\partial x^{\ell}_j}\Psi'(\bu)|<C
\ee
holds for all $\bu\in  B_{\alpha_0}$, $\ell\in \mathbb{Z}^+$, $j=1,2,3$, and  $\eta\geq 0$.
Consider $F_i^{\alpha}: V\to\mathbb{R}$, $i=1,2,3$, $\alpha\geq 0$ defined by
\be
\label{3F1}
F_1^{\alpha}(\bu)=((I+\alpha^2A)^{-1}\Psi'(\bu),\mathbf{f}),
\ee

\be
\label{3F2}
F_2^{\alpha}(\bu):=\Vinner{(I+\alpha^2A)^{-1}\Psi'(\bu),\bu}=(A(I+\alpha^2A)^{-1}\Psi'(\bu),\bu),
\ee
and
\be
\label{3F3}
F_3^{\alpha}(\bu)=\langle B(\bu,\bu),(I+\alpha^2A)^{-1}\Psi'(\bu))\rangle_{V',V}.
\ee
These three maps are well defined for $\bu\in B_\alpha$, weakly continuous in $V$, and bounded in $B_\alpha$, with uniform bounds for $\alpha\in [0,\alpha_0]$.
\end{lemma}

\noindent{\bf{Proof.}}
We first check the bound \eqref{3testboundt}. Since $\psi$ is of class $C^1$, there exists $C>0$ such that for all $\bu\in B_{\alpha_0}$, we have
\be
\left|\partial_j\psi\left((\bu,\mathbf{g}_{1}),\ldots,(\bu,\mathbf{g}_{m})\right)\right|\leq C.
\la{psibound1}
\ee
Moreover, because for every vector field $\mathbf{g}\in\mathcal{V}$, we have that
\be
\frac{\partial^{\ell}}{\partial x^{\ell}_j}\mathbf{g}\in \bigcap_{\eta\in\mathbb{Z}^+}D(A^\eta),
\la{gcap}
 \ee
 and $|A^\eta\frac{\partial^{\ell}}{\partial x^{\ell}_j}\mathbf{g}|<C$, for some $C>0$, for every $\ell\in \mathbb{Z}^+$, and $j=1,2,3$. Therefore, we obtain \eqref{3testboundt} by simple inspection of \eqref{ftest1}, \eqref{psibound1} and \eqref{gcap}.

 Now, we will prove that the maps $F_i^{\alpha}(\bu)$ are weakly
 continuous in $V$, and uniformly bounded with respect
 to $\alpha\in[0,\alpha_0]$ and $\bu\in B_{\alpha_0}$.

  Throughout this proof, let $\left\{\bu_j\right\} \subset B_{\alpha_0}$
  denote a sequence converging weakly in $V$ to a vector
  field $\bu$ in $V$. Of course, this also implies that $\bu_j$
  converges weakly in $H$ to $\bu$. By now, it is easy to see by inspection of \eqref{ftest1}, and by using \eqref{unibound} and \eqref{psibound1}, that the weak convergence in $H$ of $\bu_j$ to $\bu$ implies that
  \be
  (I+\alpha^2A)^{-1}\frac{\partial^k}{\partial x_k}A^\eta\Psi'(\bu_j)\to (I+\alpha^2A)^{-1}\frac{\partial^k}{\partial x_k}A^\eta\Psi'(\bu)
  \la{malconv1}
  \ee
  strongly in $H$, as $j\to\infty$, for every $k\in \mathbb{Z}^+$,
  and $\eta \geq 0$.
Now, because the sequence of real numbers given by the inner product in $H$ of a strongly convergent sequence in $H$, and a weakly convergent sequence in $H$  is convergent, we have
\be
F_1^{\alpha}(\bu_j)=((I+\alpha^2A)^{-1}\Psi'(\bu_j),\mathbf{f})\to ((I+\alpha^2A)^{-1}\Psi'(\bu),\mathbf{f})=F_1^\alpha(\bu),
\ee
and
\be
F_2^{\alpha}(\bu_j)=((I+\alpha^2A)^{-1}A\Psi'(\bu_j), \bu_j)\to ((I+\alpha^2A)^{-1}A\Psi'(\bu), \bu)=F_2^\alpha(\bu),
\ee
as $j\to\infty$.  This shows that $F_1^{\alpha}$ and $F_2^{\alpha}$ are weakly continuous in $V$. The uniform bounds follow by \eqref{3testboundt}, and by inspection of \eqref{3F1}.

  Now, concerning the map $F_3^\alpha$, it is easy to see by the strong convergence in $H$ of $\bu_j$ to $\bu$ established in Lemma \ref{hinv}, and by the strong convergence in $H$ stated in \eqref{malconv1}, that
  \be
B(\bu_j,(I+\alpha^2A)^{-1}\Psi'(\bu_j))\to B(\bu,(I+\alpha^2A)^{-1}\Psi'(\bu))
\la{ultconv}
\ee
strongly in $H$, as $j\to \infty$. Hence, by the anti-symmetry of the trilinear term, \eqref{asym}, by the strong convergence of $\bu_j$ to $\bu$, and by \eqref{ultconv},  it is easy to see that
  \be
\begin{aligned}
&F_3^{\alpha}(\bu_j)=-(\bu_j,B(\bu_j,(I+\alpha^2A)^{-1}\Psi'(\bu_j)))\\
&\to -(\bu,B(\bu,(I+\alpha^2A)^{-1}\Psi'(\bu)))=F_3^{\alpha}(\bu),
\end{aligned}
\ee
 as $j\to \infty$. This shows that $F_3^{\alpha}$ is weakly continuous in $V$. The uniform bounds follow from the following inequality, which is an easy consequence of \eqref{asym},
\be
\la{3nlnineq}
\left|b(\bu,\bu,\bv)\right|\leq C \left|\bu\right|^2\left|\nabla\bv\right|_{L^\infty},
\ee
for every $\bu \in V$, and $\bv\in W^{1,\infty}(\Omega)$, and from the following Sobolev's imbedding  inequality in $3$D:
\be
\la{maleta2}
\left|\varphi\right|_{L^\infty}\leq C\left|A\varphi\right|,\quad\varphi\in D(A).
\ee
Indeed, because $\partial_{x_j}\Psi'(\bu)\in D(A)$, for all $\bu \in B_{\alpha_0}$, and $j\in\left\{1,2,3\right\}$ we have  by \eqref{3testboundt}, and by \eqref{unibound}, that
\be
\begin{aligned}
|F_3^\alpha(\bu)| & \leq C\left|\bu\right|^2\left|(I+\alpha^2A)^{-1}\nabla_x\Psi'(\bu)\right|_{L^\infty}\\
&=  C\sup_{j\in\left\{1,2,3\right\}}|(I+\alpha^2A)^{-1}\frac{\partial}{\partial x_j}\Psi'(\bu)|_{L^\infty} \left|\bu\right|^2 \\
&\leq C\sup_{j\in\left\{1,2,3\right\}}|(I+\alpha^2A)^{-1}A\frac{\partial}{\partial x_j}\Psi'(\bu)|\left|\bu\right|^2\\
&   \leq C\sup_{j\in\left\{1,2,3\right\}}|A\frac{\partial}{\partial x_j}\Psi'(\bu)|\left|\bu\right|^2\leq C_{\alpha_0} \left|\bu\right|^2,
\end{aligned}
\ee
where $C_{\alpha_0}$ is uniform for $\alpha\in [0,\alpha_0]$. This implies the uniform boundedness for $F_3^\alpha$ for $\bu\in B_{\alpha_0}.\quad\Box$

\medskip

\begin{lemma}
\la{3suptest}
Let $\Psi\in\mathcal{T}_1$. Consider $F_i$ defined as in \eqref{3F1}, \eqref{3F2} and \eqref{3F3} with this choice of $\Psi$. Then, for any sequence $\left\{\mu^{\alpha_n}\right\}$ of probability invariant measures of the NSV, with $\alpha_n\rightarrow 0$, as $n\to \infty$, we have
 \be
 \label{3eigA}
 \left|\int_V \left( F_i^{\alpha_n}(\bu)-F_i^0(\bu)\right)d\mu^{\alpha_n}(\bu)\right|\to 0,\quad\text{as}\quad \alpha_n\to 0,\,\mbox{ for }\, i=1,2,3.
 \ee
\end{lemma}

\noindent{\bf{Proof.}}
Fix $\alpha_0> 0$, and now notice that because $\mu^{\alpha_n}$ are stationary statistical solutions of the NSV model, by Proposition \ref{equivalence}, we have by \eqref{boundsupp6cor} that the supports of $\mu^{\alpha_n}$ are included in $B_{\alpha_0}$, for every $\alpha_n\leq \alpha_0$. From now on, we restrict ourselves to $0<\alpha_n\leq \alpha_0$.

Now, by  inspection of \eqref{3F1}, by \eqref{opconv}, and by \eqref{3testboundt}, it is easy to see that for every $\bu\in B_{\alpha_0}$, there exists $C_{\alpha_0}>0$, such that
\be
\la{Aineq1}
\begin{aligned}
&|F_1^{\alpha_n}(\bu)-F_1^0(\bu)|  \leq C\left|\mathbf{f}\right|\left|((I+\alpha_n^2A)^{-1}-I)\Psi'(\bu)\right|\\
&\leq C\left|((I+\alpha_n^2A)^{-1}-I)\Psi'(\bu)\right|\leq  C\lambda_1^{1-\eta}\alpha_n^2 |A^{\eta}\Psi'(\bu)|\leq C_{\alpha_0} \alpha_n^2.
\end{aligned}
\ee
Now, by  inspection of \eqref{3F2}, by \eqref{opconv}, and by \eqref{3testboundt}, it is easy to see that for every $\bu\in B_{\alpha_0}$, there exists $C_{\alpha_0}>0$, such that
\be
\la{Aineq2}
\begin{aligned}
&|F_2^{\alpha_n}(\bu)-F_2^0(\bu)|  \leq C\left|\mathbf{u}\right|\left|((I+\alpha_n^2A)^{-1}-I)A\Psi'(\bu)\right|\\
&\leq C_{\alpha_0}\left|((I+\alpha_n^2A)^{-1}-I)A\Psi'(\bu)\right|\leq  C\lambda_1^{1-\eta}\alpha_n^2 |A^{\eta+1}\Psi'(\bu)|\leq C_{\alpha_0} \alpha_n^2.
\end{aligned}
\ee
For $F_3^\alpha$, because $\partial_{x_j}\Psi'(\bu)\in D(A)$, for all $\bu \in B_{\alpha_0}$, and $j\in\left\{1,2,3\right\}$, we have by \eqref{asym}, \eqref{opconv}, \eqref{3testboundt} and \eqref{maleta2} that for $\alpha_n\in [0,\alpha_0]$,
\be
\la{Aineq3}
\begin{aligned}
&|F_3^{\alpha_n}(\bu)-F_3^0(\bu)|\leq|(\bu,B(\bu,((I+\alpha_n^2A)^{-1}-I)\Psi'(\bu)))|\\
&\leq C\left|\bu\right|^2\left|((I+\alpha_n^2A)^{-1}-I)\nabla_x\Psi'(\bu)\right|_{L^\infty}\\
&=\leq C|\bu|^2\sup_{j\in\left\{1,2,3\right\}}\left|((I+\alpha_n^2A)^{-1}-I)\partial_{x_j}\Psi'(\bu)\right|_{L^\infty}\\
&\leq C_{\alpha_0}\sup_{j\in\left\{1,2,3\right\}}\left|((I+\alpha_n^2A)^{-1}-I)A\frac{\partial}{\partial x_j} \Psi'(\bu)\right|\\
&\leq C_{\alpha_0}\alpha_n^2\lambda_1^{1-\eta}\sup_{j\in\left\{1,2,3\right\}}\left|A^{\eta+1}\frac{\partial}{\partial x_j} \Psi'(\bu)\right|\leq C_{\alpha_0}\alpha_n^2.
\end{aligned}
\ee
Therefore, by \eqref{Aineq1}, \eqref{Aineq2}, \eqref{Aineq3}, and since $\mu^\nu$ are probability measures, we have
\be
\int_V |F_i^{\alpha_n}(\bu)-F_i^0(\bu)|d\mu^{\alpha_n}(\bu)\leq C_{\alpha_0}\alpha_n^2.
\ee
The result follows as $n\to\infty$.$\quad\Box$

\bigskip

Now, we prove the main result of this work.

\begin{thm}
\la{3dconv}
Given a sequence of invariant probability measures of the $3$D Navier-Stokes-Voigt model, $\left\{\mu^{\alpha_n}\right\}$, with $\alpha_n\rightarrow 0$, as $n\to \infty$, there exists a subsequence, denoted  also by $\left\{\mu^{\alpha_n}\right\}$, and a Borel probability measure $\mu$ on $V$, such that
\begin{equation}
\label{3dintlimit}
\lim_{n\to \infty}\int_{V}\Phi(\bu)d\mu^{\alpha_n}(\bu)= \int_{V}\Phi(\bu)d\mu(\bu),
\end{equation}
for all $\Phi:V\to \mathbb{R}$ weakly continuous bounded real-valued functionals. Furthermore, the weak limit measure $\mu$ is a strong stationary statistical solution of the $3$D Navier-Stokes equations.
\end{thm}
\noindent{\bf{Proof.}}
Because  $\left\{\mu^{\alpha_n}\right\}$ are invariant measures, by Proposition \ref{testethm}, we know that they are also stationary statistical solutions of the NSV model, and, therefore,
they satisfy Conditions 1, 2 and 3 in Definition \ref{nsvort}. We prove first that the sequence $\left\{\mu^{\alpha_n}\right\}$ satisfies the
tightness condition \eqref{Vtight}. Rewriting more explicitly inequality \eqref{suppargument}, we obtain
\begin{equation}
\la{3suppargument}
\int_{E_1\leq \left| \bu\right|^2+\alpha_n^2\left\|\bu\right\|^2< E_2}\left\| \bu\right\|^2d\mu^{\alpha_n}(\bu)\leq\frac{\left|\mathbf{f}
\right|^2}{\nu^2\lambda_1},
\end{equation}
for every $E_1, E_2>0$. Now, fixed  $R>0$, choose $E_1^{(n)}=\alpha_n^2R^2$, and let $E_2\to\infty$, which is possible because $\supp\mu^{\alpha_n}$ is bounded in $V$. This yields
\begin{equation}
\la{intinc}
\int_{R^2\leq \left\|\bu\right\|^2}\left\| \bu\right\|^2 d\mu^{\alpha_n}(\bu)\leq\int_{\left\{\alpha_n^2 R^2\leq \left| \bu\right|^2+\alpha_n^2\left\|\bu\right\|^2\right\}}\left\| \bu\right\|^2d\mu^{\alpha_n}(\bu)\leq\frac{\left|\mathbf{f}
\right|^2}{\nu^2\lambda_1}.
\end{equation}
The first inequality above holds because of the relation
\[
\left\{\bu\in V;\;\left\|\bu\right\|^2\geq R^2 \right\}\subset\left\{\bu\in V;\;\left|\bu\right|^2+\alpha_n^2\left\|\bu\right\|^2\geq\alpha_n^2R^2 \right\},
\]
and the fact that $\left\|\bu\right\|^2$ is positive. Therefore, denoting by $B_{R^2}$ the closed ball with radius $R^2$ in $V$, we have by \eqref{intinc} that
\[
\mu^{\alpha_n}(V\setminus B_{R^2})\leq \frac{1}{R^2}\frac{\left|\mathbf{f}
\right|^2}{\nu^2\lambda_1}.
\]
Because the closed balls $B_{R^2}$ are weakly compact in $V$, this yields the
tightness condition \eqref{Vtight}, necessary for Theorem \ref{absbor}. Therefore, it is easy to see that Theorem \ref{absbor} applies to the sequence $\mu^{\alpha_n}$, yielding the existence of a subsequence, denoted  also by $\left\{\mu^{\alpha_n}\right\}$, and a Borel probability measure $\mu$ on $V$, such that \eqref{3dintlimit} holds.

Now, we prove that $\mu$ is indeed a strong stationary statistical
solution of the NSE.  Condition $1$ of Definition \ref{3sssvort}
follows directly from \eqref{3suppargument}, and by Fatou's Lemma
\[
\int_{V}\left\| \bu\right\|^2d\mu(\bu)\leq\liminf_{n\to\infty}\int_{V}\left\| \bu\right\|^2d\mu^{\alpha_n}(\bu)\leq\frac{\left|\mathbf{f}
\right|^2}{\nu^2\lambda_1}.
\]
The proof that Condition $2$ of Definition \ref{3sssvort} is valid
follows directly from Lemma \ref{lemma2} and Lemma \ref{3suptest}.
Indeed, let us assume first that $\Psi\in \mathcal{T}_1\subset
\mathcal{T}$. Then, by Lemma \ref{lemma2}, the functions $F_i^0$ are
weakly continuous in $V$, and uniformly bounded for
$\bu\in\supp\mu^\alpha$, $\alpha\in [0,\alpha_0]$. Therefore, \be
\lim_{n\to
\infty}\int_{V}F_i^0(\bu)d\mu^{\alpha_n}(\bu)=\int_{V}F_i^0(\bu)d\mu(\bu).
\ee
Now, because the functions $F_i^{\alpha_n}$ satisfy the conditions of Lemma \ref{3suptest}, we have
\be
\la{longeps}
\begin{aligned}
0&=\lim_{n\to \infty}\int_{V}((I+\alpha_n^2A) \Psi'(\bu),f+\nu A\bu -B(\bu,\bu)) d\mu^{\alpha_n}(\bu)\\
&=\lim_{n\to \infty}\left(\int_{V} (F_1^{\alpha_n}-F_1^0)(\bu)+(F_2^{\alpha_n}-F_2^0)(\bu)+(F_3^{\alpha_n}-F_3^0)(\bu) d\mu^{\alpha_n}(\bu)\right.\\ &\left.+\int_{V} F_1^0(\bu)+F_2^0(\bu)+F_3^0(\bu) d\mu^{\alpha_n}(\bu)\right)=\int_{V} F_1^0(\bu) + F_2^0 (\bu)+ F_3^0(\bu) d\mu(\bu)\\
&=\int_{V}\langle \Psi'(\bu),f+\nu A\bu -B(\bu,\bu)\rangle_{V,V'} d\mu(\bu).
\end{aligned}
\ee

Now, we are ready to remove the extra hypothesis that assumed $\Psi\in\mathcal{T}_1\subset\mathcal{T}$. Let $\tilde\Psi\in\mathcal{T}$ be defined as in \eqref{3Itest} with the function $\tilde\psi$, and the vector fields  $\mathbf{\tilde g}_i\in V$, instead of $\psi$ and $\bg_i$. Because $\mathcal{V}$ is dense in $V$, we have that for each $\tilde \bg_i$, we can find a sequence $\left\{\bg^{(i)}_{j}\right\}_j\in \mathcal{V}$, so that $\bg^{(i)}_{j}\to \tilde \bg_i$, strongly in V, as $j\to \infty$. Now, let us define $\tilde\Psi_{j}\in \mathcal{T}_1\subset\mathcal{T}$ by simply replacing the vector fields $\tilde \bg_i$ in the definition of $\tilde\Psi$ by $\bg^{(i)}_{j}$, i.e.
\be
\label{eps3Itest}
\tilde\Psi_{j}(\bu)=\tilde\psi\left(( \bu,\bg^{(1)}_{j}),\ldots,( \bu,\bg^{(m)}_{j})\right)\in \mathcal{T}_1.
\ee

We also define $(\tilde F_i^0)_{j}$ as in \eqref{3F1}, \eqref{3F2} and \eqref{3F3}, with $\tilde \Psi$ replaced by $\tilde\Psi_{j}$. Notice that \eqref{longeps} holds for $\tilde\Psi_{j}$. It is easy to check that for every $\bu\in V$ fixed, the following convergence holds:
\be
\tilde\Psi'(\bu)_{j} \to \tilde\Psi'(\bu)\quad\mbox{ strongly in }V,\quad\mbox{as }\;j\to \infty,
\ee
which implies
\be
\la{epsilonmol}
(\tilde F _i^0)_{j}(\bu) \to \tilde F_i^0(\bu)\quad\forall\bu\in V,\quad\mbox{as }\;j\to \infty.
\ee
Moreover, it is easy to see that the sequence $\left\{\bg^{(i)}_{j}\right\}_j$ can be chosen so that there exists $C>0$ such that
\[
\left|(\tilde F_1^0)_{j}\right|+\left|(\tilde F_2^0)_{j}\right|+\left|(\tilde F_3^0)_{j}\right| \leq C\left\|\Psi'(\bu)\right\|(|\mathbf{f}|+\nu\left\|\bu\right\|+\left\|\bu\right\|^2),
\]
uniformly in $j\in\mathbb{Z}^+$. Now, because the right-hand side of the above expression is integrable, by \eqref{3dssv1}, we can use \eqref{epsilonmol} and the Lebesgue Dominated Convergence Theorem to take the limit $j\to \infty$ in \eqref{longeps}, yielding
\be
\begin{aligned}
0&=\lim_{j\to \infty}\int_{V} (\tilde F_1^0)_{j}+(\tilde F_1^0)_{j}+(\tilde F_3^0)_{j}\, d\mu(\bu)\\
&=\int_{V}\langle \tilde \Psi'(\bu),f+\nu A\bu -B(\bu,\bu)\rangle_{V,V'}\, d\mu(\bu).
\end{aligned}
\la{cond2proof}
\ee
This proves that Condition $2$ of Definition \ref{3sssvort} holds for every $\tilde\Psi\in \mathcal{T}$.

\smallskip

Now, we prove that Condition $3$ holds. Let us define
\[
\Gamma=\left\{\bu\in V; E_1\leq\left|\bu\right|^2< E_2\right\},
\]
and $\psi^{(m)}\in C^\infty(\mathbb{R}^m)$ as
\[
\psi^{(m)}(a_1,\ldots,a_m):=\sum_{j=1}^m|a_j|^2.
\]
Let us also consider a real-valued function $\chi\in C^\infty(\mathbb{R})$, and let $\left\{\mathbf{w}_j\right\}\subset (C^\infty(\Omega))^3\cap V$ be the orthonormal basis of $H$ composed of the eigenvectors of the Stokes operator, $A$. Thus, for fixed $m$, we define for $\bu\in B_{\alpha_0}$
\[
\Psi_m^{\chi}(\bu):=\frac{1}{2}\chi(\psi^{(m)}((\bu,\mathbf{w}_1),\ldots,(\bu,\mathbf{w}_m))).
\]
It is easy to see that $\Psi_m^{\chi}\in \mathcal{T}$, and that
\be
\la{testdiff}
(\Psi_m^{\chi})'(\bu)=\chi'(\sum_{j=1}^m|(\bu,\mathbf{w}_j)|^2)\sum_{j=1}^m(\bu,\mathbf{w}_j)\mathbf{w}_j.
\ee
Now, because $\chi\in C^\infty(\mathbb{R})$, we have by the Parseval identity that for every $\bu, \mathbf{h}\in H$, the following limit holds
\be
\la{testdiffonf}
\lim_{m\to\infty}((\Psi_m^{\chi})'(\bu),\mathbf{h})
=\lim_{m\to\infty}\left[\chi'(\sum_{j=1}^m|(\bu,\mathbf{w}_j)|^2)\sum_{j=1}^m(\bu,\mathbf{w}_j)(\mathbf{h},\mathbf{w}_j)\right]=\chi'(|\mathbf{u}|^2)(\bu,\mathbf{h}).
\ee
Notice that it follows from the definition of weak convergence that $\supp\mu\subset B_{\alpha_0}$,  which is bounded in $H$. Therefore, because $\mathbf{f}\in H$, we have that $((\Psi_m^{\chi})'(\bu),\mathbf{f})$ is uniformly bounded  for all $\bu\in \supp\mu$, and $m\in \mathbb{Z}^+$. Thus,  we may apply the Lebesgue dominated convergence theorem to obtain
\be
\la{f1aux}
\lim_{m\to\infty}\int_V((\Psi_m^{\chi})'(\bu),\mathbf{f})\, d\mu(\bu)=\int_V \chi'(|\mathbf{u}|^2)(\bu,\mathbf{f})\, d\mu(\bu).
\ee
Now, let us denote by $\mathbf{1}_{[E_1,E_2]}:\mathbb{R}\to\mathbb{R}$ the characteristic step function on the interval of real numbers $[E_1,E_2]\subset \mathbb{R}$, i.e.
\begin{equation*}
\mathbf{1}_{[E_1,E_2]}(x) = \left\{
\begin{array}{rl}
1 & \text{if } x \in [E_1,E_2],\\
0 & \text{otherwise}.\\
\end{array} \right.
\end{equation*}
 Consider a sequence of smooth real-valued functions $\chi_i\in C^\infty(\mathbb{R})$, so that $\chi_i'(x)\to \mathbf{1}_{[E_1,E_2]}(x)$ pointwise, as $i\to\infty$, with $|\chi_i(y)|+|\chi_i'(y)|\leq C$, for every $i\in\mathbb{Z}^+$. Thus, we may again apply the Lebesgue dominated convergence theorem to deduce that
\be
\label{3Afin1}
\lim_{i\to\infty}\lim_{m\to\infty}\int_V((\Psi_m^{\chi_i})'(\bu),\mathbf{f})\, d\mu(\bu)=\int_\Gamma (\bu,\mathbf{f})\, d\mu(\bu).
\ee
Now, we investigate the limits for the $F_2^0$ term. We obtain for every $\bu\in\supp\mu\subset V$,
\be
\begin{aligned}
& \Vinner{(\Psi_m^{\chi_i})'(\bu),\bu}=\chi_i'(\sum_{j=1}^m|(\bu,\mathbf{w}_j)|^2)\sum_{j=1}^m(\bu,\mathbf{w}_j)\Vinner{\mathbf{u},\mathbf{w}_j}\\
& =\chi_i'(\sum_{j=1}^m|(\bu,\mathbf{w}_j)|^2)\sum_{j=1}^m(\bu,\mathbf{w}_j)(\bu,A\mathbf{w}_j)=\chi_i'(\sum_{j=1}^m|(\bu,\mathbf{w}_j)|^2)\sum_{j=1}^m\lambda_j|(\mathbf{u},\mathbf{w}_j)|^2.
\end{aligned}
\ee
Therefore, because $\supp\mu\subset V$, we have
\be
\lim_{m\to \infty}\Vinner{(\Psi_m^{\chi_i})'(\bu),\bu}=\lim_{m\to\infty}\chi_i'(\sum_{j=1}^m|(\bu,\mathbf{w}_j)|^2)\sum_{j=1}^m\lambda_j|(\mathbf{u},\mathbf{w}_j)|^2
=\chi_i'(|\bu|^2)\left\|\bu\right\|^2,
\ee
for every $\bu\in \supp\mu$. Hence, by Fatou's lemma:
\be
\liminf_{m\to \infty}\int_V\Vinner{(\Psi_m^{\chi_i})'(\bu),\bu}d\mu(\bu)\geq \int_V \chi_i'(|\bu|^2)\left\|\bu\right\|^2 d\mu(\bu).
\ee
 Thus, we may again apply the Lebesgue dominated convergence theorem to deduce that
\be
\label{3Afin2}
\lim_{i\to\infty}\liminf_{m\to \infty}\int_V\Vinner{(\Psi_m^{\chi_i})'(\bu),\bu}d\mu(\bu)\geq \int_\Gamma\left\|\bu\right\|^2 d\mu(\bu).
\ee
Now, we calculate the limits for the $F_3^0$ term. Now, from \eqref{bilinear},  for $\bu\in\supp\mu\subset V$, we have $B(\bu,\bu)\in V'$, and  $A^{-1/2}B(\bu,\bu)\in H$. It is easy to prove that the following identity holds
\be
\langle B(\bu,\bu), \mathbf{w}_j\rangle_{V',V}=(A^{-1/2}B(\bu,\bu),A^{1/2}\mathbf{w}_j)=\lambda_j^{1/2}(A^{-1/2}B(\bu,\bu),\mathbf{w}_j),
\ee
for every $j=1,2,\ldots$. We refer the reader to \cite{constantinfoias}, \cite{fmrt}, or \cite{temam}, for a discussion on the properties of the negative powers of the Stokes operator. Therefore,
\be
\begin{aligned}
&\langle B(\bu,\bu), (\Psi_m^{\chi_i})'(\bu)\rangle_{V',V}\\
&=\chi_i'(\sum_{j=1}^m|(\bu,\mathbf{w}_j)|^2)\sum_{j=1}^m(\bu,\mathbf{w}_j)\langle B(\bu,\bu), \mathbf{w}_j\rangle_{V',V}\\
&=\chi_i'(\sum_{j=1}^m|(\bu,\mathbf{w}_j)|^2)\sum_{j=1}^m(\bu,\mathbf{w}_j)\lambda_j^{1/2}(A^{-1/2}B(\bu,\bu),\mathbf{w}_j)\\
&=\chi_i'(\sum_{j=1}^m|(\bu,\mathbf{w}_j)|^2)\sum_{j=1}^m(\bu,A^{1/2}\mathbf{w}_j)(A^{-1/2}B(\bu,\bu),\mathbf{w}_j)\\
&=\chi_i'(\sum_{j=1}^m|(\bu,\mathbf{w}_j)|^2)\sum_{j=1}^m(A^{1/2}\bu,\mathbf{w}_j)(A^{-1/2}B(\bu,\bu),\mathbf{w}_j).
\end{aligned}
\ee
Because for every $\bu\in\supp\mu\subset V$, we have $A^{1/2}\bu\in H$, and $A^{-1/2}B(\bu,\bu)\in H$, we obtain
\be
\begin{aligned}
&\lim_{m\to\infty}\langle B(\bu,\bu),(\Psi_m^{\chi_i})'(\bu)\rangle_{V',V}\\
&=\lim_{m\to\infty}\left[\chi_i'(\sum_{j=1}^m|(\bu,\mathbf{w}_j)|^2)\sum_{j=1}^m(A^{1/2}\bu,\mathbf{w}_j)(A^{-1/2}B(\bu,\bu),\mathbf{w}_j)\right]\\
&=(A^{1/2}\bu,A^{-1/2}B(\bu,\bu))=\langle B(\bu,\bu),\bu \rangle_{V',V}=0,
\end{aligned}
\ee
for every $\bu\in \supp\mu$. Therefore, by Fatou's lemma:
\be
\begin{aligned}
\label{3Afin3}
&\liminf_{m\to \infty}\int_V \langle B(\bu,\bu),(\Psi_m^{\chi_i})'(\bu)\rangle_{V',V} d\mu(\bu)\\
&\geq\int_V \liminf_{m\to \infty}\langle B(\bu,\bu),(\Psi_m^{\chi_i})'(\bu)\rangle_{V',V} d\mu(\bu)=0.
\end{aligned}
\ee
Therefore, using \eqref{cond2proof}, \eqref{3Afin1}, \eqref{3Afin2}, and \eqref{3Afin3}, we deduce that
\be
\begin{aligned}
& \int_{\Gamma} \left\| \bu\right\|^2 d\mu(\bu)\leq\\
&\lim_{i\to\infty}\liminf_{m\to\infty}\int_{V} \Vinner{(\Psi_m^{\chi_i})'(\bu),\bu} + \langle(\Psi_m^{\chi_i})'(\bu),B(\bu,\bu)\rangle_{V,V'} d\mu(\bu)\\
&=\lim_{i\to\infty}\lim_{m\to\infty}\int_{V} ((\Psi_m^{\chi_i})'(\bu),\mathbf{f}) d\mu(\bu)= \int_{\Gamma} ( \mathbf{f},\bu) d\mu(\bu).
\end{aligned}
\la{3lenbalance}
\ee

This proves that Condition $3$ of Definition \ref{3sssvort} is valid, concluding the proof that $\mu$ is indeed a strong stationary statistical solution of the Navier-Stokes equations.$\quad\Box$

\bigskip

Now, it follows immediately from Lemma \ref{hinv} that the function $\bu\mapsto\left|\bu\right|^2$ is weakly continuous in $V$, and bounded for $\bu\in\supp\mu^\alpha$, for $\alpha\in[0,\alpha_0]$. Thus, we can prove strong convergence for the kinetic energy.

\begin{thm}
\label{3iconvergence}
Given a sequence of invariant measures of the $3$D Navier-Stokes-Voigt model, $\left\{\mu^{\alpha_n}\right\}$, with $\alpha_n\rightarrow 0$, as $n\to \infty$, there exists a subsequence, denoted  also by $\left\{\mu^{\alpha_n}\right\}$, and a Borel probability measure $\mu$ on $V$, such that
\be
\label{3strongconvH}
\lim_{n\to \infty}\int_{V}\left|\bu\right|^2d\mu^{\alpha_n}(\bu)=\int_{V}\left|\bu\right|^2d\mu(\bu).
\ee
\end{thm}

Now, we prove our result concerning the convergence of the net energy transfer.

\begin{thm}
Given a sequence of invariant measures of the $3$D Navier-Stokes-Voigt model, $\left\{\mu^{\alpha_n}\right\}$, with $\alpha_n\rightarrow 0$, as $n\to \infty$, there exists a subsequence, denoted  also by $\left\{\mu^{\alpha_n}\right\}$, and a Borel probability measure $\mu$ on $V$, such that for every finite pair of wavenumbers $\kappa'\leq\kappa''$ with $\mathbf{f}_{\kappa',\kappa''}=0$, the convergence of the net rate of energy transfer between $\kappa'$ and $\kappa''$:
\be
\la{fbaga}
\lim_{n\to 0}(\langle e_{\kappa'}(\bu)\rangle^{\alpha_n}-\langle e_{\kappa''}(\bu)\rangle^{\alpha_n})= \int_V\left(e_{\kappa'}(\bu)- e_{\kappa''}(\bu)\right)d\mu(\bu)
\ee
holds.
\end{thm}
\noindent{\bf{Proof.}}
Let $\alpha_0>0$. Because the function $\bu\mapsto \left\|\bu_{\kappa',\kappa''}\right\|^2$  is weakly  continuous in $V$ and uniformly bounded for $\bu\in B_{\alpha_0}$, and because $\alpha_n\in [0,\alpha_0]$, for $n$ sufficiently large, we can use the convergence Theorem \ref{3dconv} to deduce
\be
\begin{aligned}
&\lim_{n\to\infty}(\langle e_{\kappa'}(\bu)\rangle^{\alpha_n}-\langle e_{\kappa''}(\bu)\rangle^{\alpha_n})=\lim_{n\to\infty}\nu\langle\left\|\bu_{\kappa',\kappa''}\right\|^2\rangle^{\alpha_n}\\
&=\nu\langle\left\|\bu_{\kappa',\kappa''}\right\|^2\rangle=(\langle e_{\kappa'}(\bu)\rangle-\langle e_{\kappa''}(\bu)\rangle).\quad\Box
\end{aligned}
\ee

\vspace{1cm}
\noindent{\bf{Acknowledgment.}}\,\,\,  This work was partially supported by the NSF grant no. DMS-0708832, the ISF grant no. 120/06, and the BSF grant no. 2004271.  F.R. was also supported by the Koshland Center for Basic Research at the Weizmann Institute of Science, and by the Hausdorff Center for Mathematics.  The authors also want to thank Professor Peter Constantin for  valuable suggestions concerning Theorem \ref{3dconv}.

\end{document}